\documentclass[10pt,final,doublecolumn]{IEEEtran}
\usepackage{amsthm}
\usepackage{amsmath}
\usepackage{latexsym}
\usepackage{graphicx}
\usepackage{bbding}
\usepackage{indentfirst}
\usepackage{algorithm,algorithmic}
\usepackage{setspace}
\usepackage{float}
\usepackage{epstopdf}
\usepackage{amssymb}
\usepackage{amsfonts}
\usepackage{enumerate}
\usepackage{multicol}
\usepackage{color}
\usepackage{diagbox}
\usepackage{bm}
\usepackage{amssymb}
\usepackage{stfloats}
\usepackage{epstopdf}
\usepackage{threeparttable}
\usepackage{epstopdf}
\usepackage{threeparttable}
\usepackage{makecell}
\usepackage{cite}
\usepackage{graphicx}
\usepackage{subcaption}
\usepackage[labelsep=period]{caption} 
\usepackage[dvipsnames,table]{xcolor} 
\definecolor{mygray}{gray}{0.55}  
\usepackage{booktabs}       
\usepackage{siunitx}        

\IEEEoverridecommandlockouts
\allowdisplaybreaks[4]

\newcommand{\Si}{\mathrm{Si}}
\newcommand{\Ci}{\mathrm{Ci}}

\def\BibTeX{{\rm B\kern-.05em{\sc i\kern-.025em b}\kern-.08em
		T\kern-.1667eb\lower.7ex\hbox{E}\kern-.125emX}}
\usepackage{balance}
\begin{document}
	\title{Flexible Coupler Antenna Enhanced Wireless Communication: Modeling and Coupler Position Optimization}
\author{{Xiaodan Shao,~\IEEEmembership{Member,~IEEE},  Chuangye Shan, Yunlong Du,  Junling Li,~\IEEEmembership{Member,~IEEE}, Rui Zhang, \IEEEmembership{Fellow, IEEE}, Cheng-Xiang Wang, \IEEEmembership{Fellow, IEEE}}
	
		\thanks{X. Shao is with the Department of Electrical and Computer Engineering, University of Waterloo, Waterloo, ON N2L 3G1, Canada (e-mail: x6shao@uwaterloo.ca).}

        \thanks{C. Shan is with the Shell (Beijing) New Energy Technology Co., Ltd, China (e-mail: chuangyeshan@outlook.com).}

    \thanks{Y. Du is with the
	School of Physics, University of Electronic Science and Technology of China,
	Chengdu 611731, China.}
			
			\thanks{J. Li and C.-X. Wang are with the National Mobile Communications Research Laboratory, School of Information Science and Engineering, Southeast University, Nanjing 211189, China (e-mail: junlingli@seu.edu.cn, chxwang@seu.edu.cn).}
			
		\thanks{R. Zhang is with the Department of Electrical and Computer Engineering, National University of Singapore, Singapore 117583 (e-mail: elezhang@nus.edu.sg).}
		
	}
		
	\maketitle
	

\begin{abstract}
	This paper proposes a novel flexible coupler antenna (FCA) that translates passive coupling elements around a fixed-position active antenna to reshape the induced currents on the passive elements for radiation. A new form of mechanical beamforming is achieved by moving only the passive coupling elements while keeping the active antenna stationary. The proposed design significantly reduces the antenna and radio-frequency (RF) chain costs of conventional active array beamforming with low mechanical control complexity and energy consumption. For the purpose of exposition, we consider a point-to-point communication system with one FCA at the transmitter and one fixed antenna at the receiver. Specifically, based on multi-port circuit theory, we establish both the line-of-sight (LoS) and multipath channel models and derive the mechanical beamforming weights of the passive couplers as functions of their positions. Then, we formulate a new problem to maximize the received signal-to-noise ratio (SNR) by optimizing the positions of passive couplers at the transmitter, subject to coupler movement and 
	transmit power constraints. Solving the resulting problem is inherently difficult because coupled channel and mechanical beamforming create non-linearity in the objective function.
	To tackle this problem, we propose an efficient block-coordinate conditional gradient method to search for the best positions of all passive couplers by sequentially optimizing the position
	of each coupler with those of the other couplers fixed in an iterative manner.
 Simulation results demonstrate that the proposed system significantly outperforms benchmark schemes in terms of achievable rate, but
 with significantly reduced active antennas and RF chains. It is validated that position reconfiguration of passive couplers provides a new and cost-effective means of enhancing wireless communication performance without the need to move active antennas.
\end{abstract}


\section{Introduction}
Multiple-input multiple-output (MIMO) communication technology will remain pivotal for future sixth-generation (6G) and beyond systems, where the demand for massive connectivity, ultra-low latency, and high energy efficiency drives the exploration of advanced wireless communication techniques \cite{7400949,10054381,Larsson2014Massive}. For instance, by
deploying more antennas than existing massive MIMO
within a given space, extremely large-scale MIMO can
achieve significantly higher achievable rates \cite{exl}. Passive MIMO,
also known as intelligent reflecting surfaces (IRS)/reconfigurable intelligent surface (RIS), can cost-effectively enhance wireless network capacity \cite{9140329,9724202,10555049,10740590}. An appealing property of MIMO is that its performance gains scale with the number of antennas. Traditional MIMO systems, however, face two main challenges. First, the higher bitrate comes at the expense of larger space requirements since more antennas are deployed, and traditional MIMO requires each antenna to be fed by a unique radio-frequency (RF) chain. This significantly increases the hardware cost of the transceiver and makes MIMO systems costly and bulky \cite{bjornson2019massive}. The associated baseband processing and RF circuitry also lead to higher power consumption, which accelerates battery depletion of wireless devices.
Second, since the antennas in conventional MIMO systems are deployed at fixed positions, the system lacks the ability to adaptively reconfigure or optimize the wireless channel within the transmit-receive region according to user distribution and environmental variations \cite{9903389,11278723}. To fully exploit spatial channel variations, the six-dimensional movable antenna (6DMA) has recently been proposed to increase the MIMO system capacity without requiring additional antennas \cite{shao20246d,6dma_dis,10945745}. This technique utilizes the adaptability of antenna positions and rotations (that is, orientations) in three-dimensional (3D) space at transceivers to allocate antenna resources according to the spatial distribution of channels. Several efficient position- and rotation-optimization algorithms and channel-estimation algorithms have been studied in \cite{jiang2025statistical,shao2025tutorial,liu2024uav,6DMA_JSTSP,li2025ai}. Despite its promising potential, existing designs of 6DMA move the active antenna element together with its RF feed over a sufficiently large region of size in the order of several to tens of signal wavelengths to exploit spatial channel variations, which introduces substantial mechanical complexity and energy consumption.
\begin{figure*}[t!]
	\centering
	\setlength{\abovecaptionskip}{0.cm}
	\includegraphics[width=6.3in]{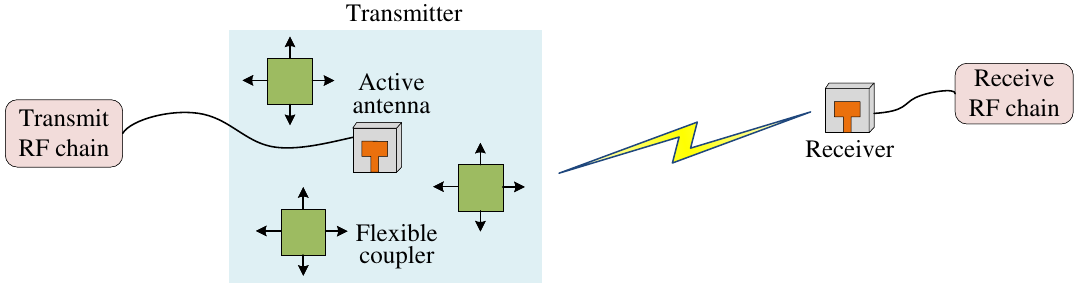}
	\caption{Flexible coupler antenna-aided point-to-point communication system.}
	\label{ca}
\end{figure*}

To construct a flexible and low-cost antenna system that can still achieve most of the capacity gains of state-of-the-art MIMO systems, we propose in this paper a new flexible coupler antenna (FCA) as shown in Fig.~\ref{ca}. In the FCA system, a set of flexible passive couplers are arranged around an active antenna at the transceiver. The passive couplers can be independently adjusted in terms of their positions, while the active antenna is connected to an RF chain at the transceiver and remains fixed. 
Mutual coupling, which refers to the electromagnetic (EM) interaction between antennas, has traditionally been regarded as a drawback in antenna array design because coupled power may be fed back into the circuit as a reflected wave, thereby reducing antenna efficiency \cite{7945283, 10500503,10971994}. In contrast, the proposed FCA exploits mutual coupling as an enabler of advanced spatial processing. Specifically, by exploiting the coupler movement and the mutual coupling between neighboring coupler and antenna elements, the passive couplers radiate through the excitation induced by the active antenna, enabling the system to perform a new form of position-dependent mechanical beamforming.
Therefore, the transceiver can design and control the coupler positions according to the spatial channel distribution to maximize the system capacity (see Fig.~\ref{ca}).  
The proposed FCA-based communication system requires the integration of a coupler positioning module together with the conventional communication module at the transceiver \cite{Shao2026FC,Shao2026FCmaga,shao2026rotatable,FCjstsp}. One feasible approach is to employ micro-electromechanical systems (MEMS) to facilitate the movement of the couplers within the designated region. Since miniaturized MEMS devices provide high performance and very low power consumption with electrostatic actuation \cite{balanis2011modern}, a MEMS-enabled FCA offers high positioning accuracy, making it suitable for small-scale coupler systems on the order of the wavelength. This method enables precise control of the coupler positions through a dedicated central processing unit (CPU) to meet communication requirements. The compact structure of the flexible coupler antenna makes it particularly attractive for devices with stringent size, weight, and power (SWAP) constraints.

It is worth noting that the FCA system proposed in this paper differs significantly from the existing 6DMA system \cite{shao20246d,6dma_dis} and the conventional electronically steerable parasitic array radiator (ESPAR) system \cite{8628984,ESP}.
First, compared with the 6DMA that enhances source-destination transmission by translating and/or rotating the active antennas with RF chains \cite{shao2025tutorial}, the FCA operates as a passive array without additional RF chains. It generates coupling signals through EM interaction and moves only the passive couplers within a small region, without the need to move the active antenna or its RF chain. Hence, it requires lower transmit-power consumption and reduced mechanical-control complexity.
Furthermore, the movement region of the flexible coupler is on the order of a few wavelengths to maintain strong mutual coupling, whereas 6DMA operates over much larger regions. 
This makes the movement of flexible couplers faster and easier to implement in practice.
From an implementation viewpoint, compared with 6DMA, the FCA has advantages such as compact structure, low profile, and light weight, making it suitable for small-sized terminals and practical deployment. For instance, when FCAs are integrated into mobile devices, the received signal strength at the user side can be significantly enhanced without the need to deploy additional antennas. 
Second, unlike traditional ESPAR systems whose performance is generally inferior to that of fully active adaptive arrays, the FCA-based communication system can surpass fully active arrays in achievable performance. This is because an ESPAR antenna consists of a single driven element surrounded by passive elements connected to reactive loads, and its pattern can be reconfigured by varying the reactance values \cite{ESP}, yet the finite adjustment range and the series resistance of varactors introduce loss and mismatch. In contrast, the FCA enhances communication performance by optimizing coupler positions to obtain mechanical-beamforming gain, and by further exploiting spatial multiplexing and geometric gains within the movement region while effectively suppressing interference. These characteristics make the FCA a promising technology for future wireless networks, particularly for terminals with limited installation space. Note that, unlike ambient backscatter communication \cite{7551180}, the proposed FCA does not rely on backscatter modulation for information transmission. Instead, its passive couplers assist the active antenna through position-dependent mutual coupling.
 A summary of the above
comparison among the proposed  FCA and existing 6DMA
as well as traditional ESPAR is given in Table I, where $N$ denotes the number of active antennas in 6DMA.
\begin{table*}[!t]
	\small
	\caption{Comparison of flexible coupler antenna with other related technologies.}
	\label{Table1}
	\centering
	\begin{tabular}{|c|c|c|c|c|c|c|c|c|}
		\hline
		\makecell{Architecture} & \makecell{Tunable parameter} & \makecell{Operating mechanism} &\makecell{Number of RF \\ chains needed} &\makecell{Performance \\Gain}&\makecell{Hardware/energy\\cost}\\
		\hline
		\makecell{Flexible coupler antenna}  &\makecell{Position/rotation \\ of passive coupler}  & \makecell{Passive \\ mutual coupling} &1& High& Low\\
		\hline
		6DMA \cite{shao20246d,6dma_dis} & \makecell{\makecell{Position/rotation \\of active antenna}}   & \makecell{Active radiation}  & $N$   & Very high& Medium \\
		\hline
		\makecell{ESPAR \cite{8628984}}  & Antenna impedance  & \makecell{Passive \\ mutual coupling} &1& Low& Low\\
		\hline
	\end{tabular}
\end{table*}

The main contributions of this paper are summarized as follows:
\begin{itemize}
	\item 
	We propose a new FCA architecture that translates passive coupling elements around a fixed active antenna to enhance wireless communication performance. By employing the multi-port circuit theory framework, we establish both the line-of-sight (LoS) and multipath channel models and derive the beamforming weights of the passive couplers in terms of coupler positions in the FCA-enhanced system. Compared with the 6DMA that moves active antennas over a large spatial region, the proposed approach achieves mechanical beamforming by relocating only passive couplers within a small area through EM coupling without the need to move active antennas/RF chains, which greatly reduces mechanical control complexity and energy consumption.
	
	\item  
	Based on the established model, we formulate an optimization problem to maximize the receiver signal-to-noise ratio (SNR) by designing the coupler position vector, subject to movement and transmit power constraints. To solve this non-convex problem, we propose an efficient block-coordinate conditional gradient method that searches for the optimal positions of all passive couplers by sequentially optimizing the position of each coupler while fixing the others in an iterative manner. 
	
	\item
	Extensive simulations are conducted to validate the effectiveness of the proposed FCA architecture compared with various benchmark systems. The results show that under the LoS channel, the FCA approaches the fully active array in achievable rate by exploiting mechanical beamforming. In multipath channels, by jointly leveraging the mechanical beamforming gain and the fading-mitigation gain, the achievable rate of the proposed architecture with compact wavelength-scale movements surpasses that of fully active arrays while using fewer RF chains and lower processing power. Moreover, the achievable rate of the FCA system increases with the number of passive couplers, thereby achieving higher throughput in a cost-efficient manner.
\end{itemize}

The remainder of this paper is organized as follows. Section~II describes the channel and signal models of the FCA system. Section~III formulates the optimization problem for maximizing the receiver SNR via position optimization of the flexible couplers at the transmitter. Section~IV presents the proposed conditional-gradient-based algorithm for solving the problem. Section~V provides numerical results and corresponding discussions. Finally, Section~VI concludes the paper.

\emph{Notations.}
Boldface uppercase and lowercase letters denote matrices and vectors, respectively.
$(\cdot)^{*}$, $(\cdot)^{\mathrm H}$, and $(\cdot)^{\mathrm T}$ denote complex conjugation, conjugate transpose, and transpose, respectively.
$\mathbb{E}[\cdot]$ denotes statistical expectation.
For a scalar $a$, $|a|$ denotes its magnitude.
For a vector $\mathbf{a}$, $\|\mathbf{a}\|_2$ denotes the $\ell_{2}$ norm.
$\mathrm{diag}(\mathbf{x})$ denotes a diagonal matrix whose diagonal is $\mathbf{x}$.
$[\mathbf{a}]_{j}$ denotes the $j$th entry of a vector $\mathbf{a}$, and $[\mathbf{A}]_{i,j}$ denotes the $(i,j)$ entry of a matrix $\mathbf{A}$.
$\otimes$ denotes the Kronecker product.
$\mathbf{I}_{M}$ denotes the $M\times M$ identity matrix, and $\mathbf{1}_{N}$ denotes the $N\times1$ all-ones vector.
$\mathrm{blkdiag}\{\mathbf{A}_{1},\ldots,\mathbf{A}_{M}\}$ denotes a block diagonal matrix with diagonal sub-matrices given by $\mathbf{A}_{1},\ldots,\mathbf{A}_{M}$.
$\mathrm{vec}(\mathbf{A})$ stacks the columns of $\mathbf{A}$ into a single column vector.
$\mathcal{O}(\cdot)$ denotes the big O notation.
$\mathbb{R}$ and $\mathbb{C}$ denote the real and complex fields, respectively.
$\mathcal{CN}(\mu,\sigma^{2})$ denotes a complex Gaussian distribution with mean $\mu$ and variance $\sigma^{2}$.
$\Re\{\cdot\}$ and $\Im\{\cdot\}$ denote the real and imaginary parts of a complex quantity, respectively.
$\angle x$ denotes the phase of a complex number $x$.

\section{System Model}
As shown in Fig.~\ref{ca}, the proposed FCA single-user communication system consists of one transmitter and one receiver. At the transmitter, there are one fixed-position active antenna and $N$ passive couplers, which can move within a designated area around the active antenna. The positions of the passive couplers can be mechanically adjusted with the aid of drive components, such as MEMS with low power consumption and fast response \cite{balanis2011modern}. The active antenna is connected to a single RF chain. The receiver is assumed to have a fixed single antenna. In this configuration, transmit signals of couplers are generated through the induced currents on the passive couplers via near-field EM coupling, without requiring any additional RF feeds. Moreover, the proposed FCA can enhance the wireless communication performance by relocating only passive couplers within a small area without the need to move active antennas or RF chains, which greatly reduces mechanical control complexity and energy consumption.

\subsection{Wireless Propagation
	Channel}
\subsubsection{LoS Channel}
We first consider a far-field LoS channel model. 
We assume the fixed position of the active antenna is $\mathbf{p}_0=[0, 0, 0]^{\mathrm T}$, which is the origin of a global Cartesian coordinate system (CCS), and the center position vector of the $n$-th coupler is $\mathbf{p}_n$ $(n\in\mathcal{N}\triangleq\{1,2,\cdots,N\})$, which is given by
\begin{align}
	\mathbf{p}_n=[x_n,y_n,z_n]^{\mathrm T}\in\mathcal{C}, \label{bb1}
\end{align}
where $\mathcal{C}$ denotes the given space at the transmitter in which the couplers can be flexibly positioned. We assume that $\mathcal{C}$ is a convex set which has a finite size. In \eqref{bb1}, $x_n$, $y_n$ and $z_n$ represent the coordinates of the $n$-th coupler's center in the CCS $o\text{-}xyz$, with $z_n=0, \forall n\in \mathcal{N}$.

Let $\mathbf f(\phi) =[\cos\phi,\ \sin\phi,\ 0]^{\mathrm T}$ denote the pointing vector of the LoS path with azimuth angle $\phi\!\in\![-\pi,\pi]$. 
For the $n$-th flexible coupler at $\mathbf p_n=[x_n,y_n,0]^{\mathrm T}\in\mathcal C$, the LoS steering factor relative of couplers is
\begin{align}
	\mathbf{a}(\mathbf p)\!=\!\left[\!\exp\!\big(-\mathrm{j} \frac{2\pi}{\lambda}\,\mathbf f(\phi)^{\mathrm T}\mathbf p_1\big),\cdots,\exp\!\big(-\mathrm{j} \frac{2\pi}{\lambda}\,\mathbf f(\phi)^{\mathrm T}\mathbf p_N\big)\!\right]^{\mathrm T},
\end{align}
where $\lambda$ is the carrier wavelength. 
Stacking all ports (active antenna and passive couplers) with the active antenna indexed by $0$, we obtain the LoS channel vector as
\begin{align}
	\mathbf h(\mathbf p)=\gamma\,\big[1, \mathbf{a}^{\mathrm T}(\mathbf p)\big]^{\mathrm T}\in \mathbb{C}^{(N+1)\times 1},
	\label{eq:los-h}
\end{align}
where $\gamma$ denotes the common large-scale attenuation for all ports  and $\mathbf p=[\mathbf p_1^{\mathrm T},\mathbf p_2^{\mathrm T},\cdots,\mathbf p_N^{\mathrm T}]^{\mathrm T}$ is the position vector of all couplers. 

\subsubsection{Multipath Channel}
Next, we consider a far-field multipath channel model. 
Let $\mathbf f_l=[\cos\phi_l,\ \sin\phi_l,\ 0]^{\mathrm T}$ be the pointing vector of path $l$ with azimuth $\phi_l\!\in\![-\pi,\pi]$. Define the steering vector of the $n$-th coupler at $\mathbf p_n=[x_n,y_n,0]^{\mathrm T}$ as
\begin{align}
	\mathbf t(\mathbf p_n)=\big[e^{-\mathrm{j}\frac{2\pi}{\lambda}\mathbf f_1^{\mathrm T}\mathbf p_n},\ldots,e^{-\mathrm{j}\frac{2\pi}{\lambda}\mathbf f_L^{\mathrm T}\mathbf p_n}\big]^{\mathrm T}\!\in\!\mathbb C^{L\times1}.
\end{align}
By combining all ports, we obtain the multipath channel between the FCA and the user as
\begin{align}
	\mathbf h(\mathbf p)&=\mathbf G(\mathbf p)^{\mathrm H}\boldsymbol{\gamma}\in\!\mathbb C^{(N+1)\times1},
	\label{eq:hoc}
\end{align}
with 
\begin{align}
	\mathbf G(\mathbf p)&=\big[\,\mathbf 1_L,\ \mathbf t(\mathbf p_1),\ldots,\mathbf t(\mathbf p_N)\,\big]\in\mathbb C^{L\times(N+1)},
	\label{eq:hoc}
\end{align}
where $L$ denotes the total number of paths, and $\boldsymbol{\gamma} = [\gamma_1, \ldots, \gamma_L]^{\mathrm T} \in \mathbb{C}^{L \times 1}$ collects the complex path gains. In \eqref{eq:hoc}, the steering vector for the active antenna is denoted by $\mathbf{t}(\mathbf{p}_0) = \mathbf{1}_L$.

\subsection{Mechanical Beamforming}
Let the voltage at the port of the active antenna be denoted as $v_0$ and the current as $i_0$. 
The vector of voltages at the passive  coupler ports is denoted as $\mathbf v_{\mathrm{E}} = [v_1, \ldots, v_{N}]^{\mathrm T}\in \mathbb{C}^{N\times 1}$. 
We then define $\mathbf v_{\mathrm{TX}} = [v_0, \mathbf v_{\mathrm{E}}^{\mathrm T}]^{\mathrm T}\in \mathbb{C}^{(N+1)\times 1}$ for the voltages of all active and passive ports at the transmitter. 
Similarly, the current vector at the passive coupler ports is denoted by $\mathbf i_{\mathrm{E}} = [i_1, \ldots, i_{N}]^{\mathrm T}\in \mathbb{C}^{N\times 1}$. 
Thus, the current vector of all active and passive ports at the transmitter is given by
\begin{align}
	\mathbf i_{\mathrm{TX}} = [i_0, \mathbf i_{\mathrm{E}}^{\mathrm T}]^{\mathrm T}\in \mathbb{C}^{(N+1)\times 1}. \label{iii}
\end{align} 
Note that the first element of vectors $\mathbf i_{\mathrm{TX}}$ and $\mathbf v_{\mathrm{TX}}$ corresponds to the active antenna.
\begin{figure}[t!]
	\centering
	\setlength{\abovecaptionskip}{0.cm}
	\includegraphics[width=3.1in]{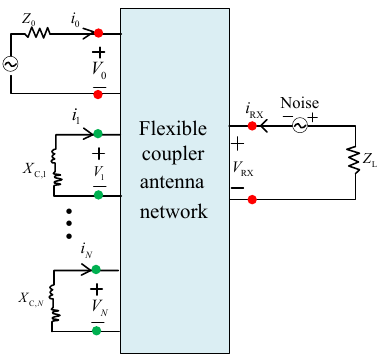}
	\caption{Circuit model of the FCA system, where only one antenna is connected to the RF chain while the remaining couplers are passive and movable.}
	\label{circuit}
\end{figure}

In
the circuit-theoretic formulation, the EM interaction between
different ports/antennas is captured in the form of an impedance
matrix. The voltage at each port is related to the
currents of all ports through the impedance matrix, denoted by $\mathbf Z_{\mathrm{TX}}(\mathbf{p})$.
Specifically, let the self-impedance of the active antenna be $z_{00}$. 
The vector of mutual impedance between the active antenna and $N$ flexible couplers is denoted by  $\bar{\mathbf z}(\mathbf{p})\in \mathbb{C}^{N\times 1}$. 
The mutual impedance matrix of the $N$ flexible couplers is denoted by  $\mathbf Z_{\mathrm{E}}(\mathbf{p})\in \mathbb{C}^{N\times N}$. Note that both $\bar{\mathbf z}(\mathbf{p})$ and $\mathbf Z_{\mathrm{E}}(\mathbf{p})$ depend on the positions of flexible couplers, $\mathbf{p}$.  
We use the block matrix notation to express $\mathbf Z_{\mathrm{TX}}(\mathbf{p}) $ as 
\begin{align}\label{bq}
	\mathbf Z_{\mathrm{TX}}(\mathbf{p}) =
	\begin{bmatrix}
		z_{00} & \bar{\mathbf z}^{\mathrm T}(\mathbf{p})\\
		\bar{\mathbf z}(\mathbf{p}) & \mathbf Z_{\mathrm{E}}(\mathbf{p})
	\end{bmatrix}\in \mathbb{C}^{(N+1)\times (N+1)}.
\end{align}

Let
the transmit information symbol be $s\in\mathbb{C}$
and we assume it has zero mean and unit variance.  We
use the antenna current vector, $\mathbf{i}_{\mathrm{TX}}$, to represent the equivalent transmit precoding/beamforming 
vector. Hence, the transmit signal is $\mathbf{i}_{\mathrm{TX}} s$.
Beamforming in an antenna array is achieved by configuring the current vector $\mathbf i_{\mathrm{TX}}$. 
In conventional active array, the magnitude and phase of each element in $\mathbf i_{\mathrm{TX}}$ can be adjusted independently, subject to a given total transmit power constraint.  
In contrast, FCA creates beam patterns using a combination of the control of the active antenna amplitude and phase as well as the mutual coupling among all active and passive elements (through adjusting $\mathbf{p}$).  

The circuit model relates the receive voltage at the receiver to the transmit
current. The impedance parameter of the single receive antenna is $Z_{\mathrm{RX}} \in \mathbb{C}$. The voltage and current at the receiver antenna port are denoted by $v_{\mathrm{RX}} \in \mathbb{C}$ and $i_{\mathrm{RX}} \in \mathbb{C}$, respectively.
In the adopted multi-port circuit model, the wireless propagation channel impedance vector can be represented by the path-based physical channel vector $\mathbf h(\mathbf p)$, which properly reflects the relation between the transmitter and the receiver \cite{4685923}.
We thus  have 
\begin{subequations}\label{eq:1}
	\begin{align}
		\mathbf{v}_{\mathrm{TX}} &= \mathbf{Z}_{\mathrm{TX}}(\mathbf{p}) \mathbf{i}_{\mathrm{TX}} s + \mathbf h(\mathbf p) i_{\mathrm{RX}}, \label{eq:1a}\\
		v_{\mathrm{RX}} &= \mathbf h(\mathbf p)^{\mathrm T} \mathbf{i}_{\mathrm{TX}} s + Z_{\mathrm{RX}} i_{\mathrm{RX}}. \label{eq:1b}
	\end{align}
\end{subequations}

To simplify the system model by decoupling $\mathbf{v}_{\mathrm{TX}}$ from
$i_{\mathrm{RX}}$, the unilateral approximation \cite{balanis2016antenna} is applied, yielding
\begin{align}
	\mathbf{v}_{\mathrm{TX}} \approx \mathbf{Z}_{\mathrm{TX}}(\mathbf{p}) \mathbf{i}_{\mathrm{TX}} s. \label{eq:unilateral}
\end{align} 
The unilateral approximation is reasonable, since the receive-side electromagnetic back-action on the transmitter is negligible in the far-field regime.
Using the approximation in \eqref{eq:unilateral}, we have
\begin{align}
	[v_0,\ \mathbf v_{\mathrm E}^{\mathrm T}]^{\mathrm T}
	=
	\mathbf Z_{\mathrm{TX}}
	[i_0s,\ \mathbf i_{\mathrm E}^{\mathrm T}s]^{\mathrm T}.
	\label{4}
\end{align}

The magnitude and phase of $i_0$ can be controlled independently through the RF chain. 
Of note, however, entries in $\mathbf i_{\mathrm{E}}$ cannot be tuned independently as the current on each flexible coupler is induced from $i_0$ through EM coupling.  

We now establish the dependence of $\mathbf i_{\mathrm{E}}$ on $i_0$. 
As shown in Fig. \ref{circuit}, $Z_L$ denotes the load impedance on the receive
antenna and the $n$-th flexible coupler is terminated with load $X_{\mathrm {C},n}$. We define the coupler load impedance matrix as
\begin{align}
	\mathbf X = \mathrm{diag}\{X_{\mathrm {C},1}, \ldots, X_{\mathrm {C},N}\}.
\end{align}
Under the standard port convention, the port voltage has the opposite sign of the port current \cite{opp}. Hence, the voltage vector at the flexible couplers is 
\begin{align}\label{pi9}
	\mathbf v_{\mathrm{E}} = -\mathbf X \mathbf i_{\mathrm{E}}s.
\end{align}
On the other hand, from \eqref{bq} and \eqref{4} we have
\begin{align}\label{pi8}
	\mathbf v_{\mathrm{E}} = \bar{\mathbf z}(\mathbf{p})i_0s +\mathbf Z_{\mathrm{E}}(\mathbf{p})\mathbf i_{\mathrm{E}}s.
\end{align}

By equating \eqref{pi9} and \eqref{pi8}, the current vector at the flexible couplers can be obtained as
\begin{subequations}
	\begin{align}
		\mathbf{i}_{\mathrm{E}}& = -(\mathbf{Z}_{\mathrm{E}}(\mathbf{p}) + \mathbf X)^{-1}\bar{\mathbf{z}} (\mathbf{p})i_0\\
		&=-\mathbf{w}(\mathbf{p})i_0,\label{eq:5}
	\end{align}
\end{subequations}
where 
\begin{align}\label{hhh}
	\mathbf{w}(\mathbf{p})\triangleq(\mathbf{Z}_{\mathrm{E}}(\mathbf{p}) + \mathbf X)^{-1}\bar{\mathbf{z}}(\mathbf{p})\in \mathbb{C}^{N\times 1},
\end{align}
is the equivalent weight vector for transmit beamforming, which we call the \textbf{mechanical beamforming} vector, since it can be adjusted through mechanically moving the positions of passive couplers, $\mathbf{p}$.

\subsection{Signal Model}
Stacking the active and flexible coupler currents yields the overall transmit current vector
\begin{align}\label{so}
	\mathbf i_{\mathrm{TX}}
	= \tilde{\mathbf w}(\mathbf p)\, i_0,
\end{align}
where 
\begin{align}
	\tilde{\mathbf w}(\mathbf p)\triangleq[1, -\mathbf w(\mathbf p)^{\mathrm T}]^{\mathrm T}\in \mathbb{C}^{(N+1)\times 1},
\end{align}
denotes the overall mechanical beamforming vector, which includes both the active antenna and the passive couplers.

Since the outputs of couplers cannot be observed, instead they are electronically
combined with each other and the active element. According to \eqref{eq:5},
the outputs of the FCA are collected from the single active port.
Substituting \(\mathbf i_{\mathrm{TX}}\) in \eqref{so} into the receive-port relation in \eqref{eq:1b} gives the observable single-RF output
\begin{align}
	\label{q9}
	y= \mathbf h^{\mathrm T}(\mathbf p)\tilde{\mathbf w}(\mathbf p) i_0s + z,
\end{align}
where \(z \sim \mathcal{CN}(0,\sigma^2)\) denotes complex additive white Gaussian noise with zero mean and variance \(\sigma^2\). In \eqref{q9}, $i_0s$ is equivalent to redefining the effective symbol amplitude and phase of \(s\).

Finally, using \eqref{q9}, the SNR at the receiver is given by
\begin{subequations}\label{buu1}
	\begin{align}
		&r(\mathbf{p})=\frac{|i_0|^2}{\sigma^2}\big| \mathbf h^{\mathrm T}(\mathbf p)\tilde{\mathbf w}(\mathbf p)\big|^2\label{buu}\\
		&=\frac{\gamma^2}{\sigma^2}|i_0|^2\left|1-\mathbf{a}^{\mathrm T}(\mathbf p)(\mathbf{Z}_{\mathrm{E}}(\mathbf{p}) + \mathbf X)^{-1}\bar{\mathbf{z}}(\mathbf{p})\right|^2.
	\end{align}
\end{subequations}
It is worth noting that,
in contrast to the conventional ESPAR channel with fixed-position couplers \cite{8628984}, the rate of the FCA-enabled
wireless channel in \eqref{buu1} is determined by coupler positions $\mathbf p$.

\section{Problem Formulation}
According to circuit theory \cite{5446312}, the total transmit power radiated from
the FCA is given by
\begin{align} \label{35}
	P_{\mathrm{TX}} = \mathbb{E}\!\left\{ \Re\!\big[(\mathbf{i}_{\mathrm{TX}} s)^{\mathrm H} \mathbf{v}_{\mathrm{TX}}\big] \right\},
\end{align}
where the expectation is over the random information symbols. Substituting $\mathbf{v}_{\mathrm{TX}} \approx \mathbf{Z}_{\mathrm{TX}} (\mathbf{p})\mathbf{i}_{\mathrm{TX}} s$ into \eqref{35} gives
\begin{align}
	P_{\mathrm{TX}}
	&= \mathbb{E}\!\left\{ \Re\!\big[s^* \mathbf{i}_{\mathrm{TX}}^{\mathrm H} \mathbf{Z}_{\mathrm{TX}}(\mathbf{p}) \mathbf{i}_{\mathrm{TX}} s \big] \right\} \notag\\
	&= \mathbf{i}_{\mathrm{TX}}^{\mathrm H} \Re\!\{\mathbf{Z}_{\mathrm{TX}}(\mathbf{p})\} \mathbf{i}_{\mathrm{TX}}.
	\label{ii1}
\end{align}

By substituting \eqref{so} into \eqref{ii1}, we obtain the transmit power constraint,
\begin{align}\label{ui0}
	|i_0|^2\, \tilde{\mathbf{w}}(\mathbf{p})^{\mathrm H}\, \Re\{\mathbf{Z}_{\mathrm{TX}}(\mathbf{p})\}\, \tilde{\mathbf{w}}(\mathbf{p}) \le P_{\max},
\end{align}
where $P_{\max}$ denotes the maximum transmit power.

Next, we aim to maximize the received SNR of the FCA wireless system by optimizing the positions $\mathbf p$ of all flexible couplers at the transmitter. For the purpose of exposition, in this paper, we assume that the channel $\mathbf{h}(\mathbf p)$ is perfectly known at the transmitter for all possible positions of the couplers, $\mathbf{p}$, in their admissible movable region $\mathcal C$. Accordingly, the optimization problem is formulated as follows:
\begin{subequations}\label{MG30}
	\begin{align}
		\text{(P1)}~
		&\max_{\{\mathbf p_n\}_{n=1}^N}~ 
		r(\mathbf{p})\\
		\text{s.t.}~~
		& \eqref{ui0},\\
		& \mathbf p_n\in \mathcal C, \forall n\in \mathcal{N}, \label{z81}\\
		& \|\mathbf p_m-\mathbf p_{m'}\|_2 \ge d_{\min},~
		\forall m,m'\in\{0\}\cup\mathcal{N},~m\neq m', \label{z82}
	\end{align}
\end{subequations}
where $d_{\min}$ represents the minimum allowable center-to-center distance between any two couplers or between a coupler and the active antenna, which ensures that physical overlap between them does not occur.

Since the objective \(r(\mathbf p)\) in (P1) is strictly monotone in \(|i_{0}|^{2}\) for fixed \(\mathbf p\) and \(\tilde{\mathbf w}(\mathbf p)\), the optimum satisfies the power budget in \eqref{ui0} with equality. Hence, the power constraint is active at the
optimum for problem (P1). In addition, the matrix \(\Re\{\mathbf Z_{\mathrm{TX}}(\mathbf p)\}\) is positive semidefinite, which ensures a nonnegative denominator and a well-defined scaling factor.
Thus, according to \eqref{ui0}, we obtain 
\begin{align} \label{ui8}
	|i_0|^2 = 
	\frac{P_{\max}}
	{\tilde{\mathbf w}^{\mathrm H}(\mathbf{p})\Re\{\mathbf Z_{\mathrm{TX}}(\mathbf{p})\}\tilde{\mathbf w}(\mathbf{p})}.
\end{align}
Substituting \eqref{ui8} into \eqref{buu1}, we have
\begin{align}\label{ffff}
	&r(\mathbf{p})=\frac{P_{\max}}{\sigma^2}
	\frac{| \mathbf h^{\mathrm T}(\mathbf p)\tilde{\mathbf w}(\mathbf p)|^2}
	{\tilde{\mathbf w}^{\mathrm H}(\mathbf{p})\Re\{\mathbf Z_{\mathrm{TX}}(\mathbf{p})\}\tilde{\mathbf w}(\mathbf{p})}.
\end{align}
Dropping the constant factors in \eqref{ffff}, problem (P1) reduces to the following  problem,
\begin{subequations}
	\begin{align}
		\text{(P2)}~
		&\max_{\{\mathbf p_n\}_{n=1}^N}~ 
		\frac{| \mathbf h^{\mathrm T}(\mathbf p)\tilde{\mathbf w}(\mathbf p)|^2}
		{\tilde{\mathbf w}^{\mathrm H}(\mathbf{p})\Re\{\mathbf Z_{\mathrm{TX}}(\mathbf{p})\}\tilde{\mathbf w}(\mathbf{p})}\\
		\text{s.t.}~~
		& \mathbf p_n\in \mathcal C, \forall n\in \mathcal{N},\label{zc81}\\
		& \|\mathbf p_m-\mathbf p_{m'}\|_2 \ge d_{\min},~
		\forall m,m'\in\{0\}\cup\mathcal{N},~m\neq m'. \label{zc82}
	\end{align}
\end{subequations}

Note that problem (P2) is a nonconvex optimization problem because the objective function is nonconcave with respect to the coupler positions $\mathbf{p}$, and the constraint in \eqref{zc82} is also nonconvex. Moreover, the coupler positions $\mathbf{p}$ are interdependent in both the channel response and the mechanical beamforming, which makes the coupler-position optimization a challenging task.

\section{Algorithm Design for Coupler Position Optimization}
To solve (P2) efficiently, we adopt a block-coordinate conditional gradient method.
At each step, we keep the positions of all couplers fixed except for $\mathbf p_n$ and optimize over it.
We define the single-variable objective function as
\begin{align}\label{kl}
	\widetilde{R}(\mathbf p_n)
	\triangleq 
	R(\mathbf p_1,\ldots,\mathbf p_{n-1},\mathbf p_n,\mathbf p_{n+1},\ldots,\mathbf p_N),
\end{align}
where $\{\mathbf p_1,\ldots,\mathbf p_{n-1},\mathbf p_{n+1},\ldots,\mathbf p_N\}$ are held fixed and only $\mathbf p_n$ is treated as a variable.
This reparameterization enables a conditional-gradient update for $\mathbf p_n$ while keeping the remaining coupler positions unchanged. The corresponding subproblem is
\begin{subequations}\label{eq:P2}
	\begin{align}
		\text{(P3)}\;
		&\max_{\mathbf p_n}\ \widetilde{R}(\mathbf p_n), \label{eq:p2b_obj_d}\\
		\text{s.t. }\;
		& \mathbf p_n\in\mathcal C, \label{ax1}\\
		& \|\mathbf p_n-\mathbf p_j\|_2 \ge d_{\min},\ \forall j\neq n, j\in \{0\}\cup\mathcal{N}. \label{ax2}
	\end{align}
\end{subequations}

In the above, constraint \eqref{ax1} is convex, while the objective of (P3) is nonconcave and constraint \eqref{ax2} is nonconvex with respect to $\mathbf p_n$.
Thus, it is difficult to obtain a globally optimal solution for (P3) efficiently.
We first transform the nonconvex constraint in \eqref{ax2} into a convex form and then apply a feasible-direction method, namely the conditional gradient method \cite{linear}, to solve this problem.

We linearize the nonconvex constraint in \eqref{ax2} as follows.
At iteration $t$, with current point $\mathbf p_n^{(t-1)}$ and for each coupler/antenna $j\neq n$, we define the boundary point obtained by intersecting the vector $\mathbf p_n^{(t-1)}-\mathbf p_j$ with the circle $\|\mathbf p_n-\mathbf p_j\|_2=d_{\min}$:
\begin{align}\label{eq:bdry_d}
	\mathbf p_{\mathrm{bdry},j}
	= \mathbf p_j - \frac{d_{\min}}{\|\mathbf p_j-\mathbf p_n^{(t-1)}\|_2}
	\big(\mathbf p_j-\mathbf p_n^{(t-1)}\big).
\end{align}
A supporting halfspace with respect to $\mathbf p_{\mathrm{bdry},j}$ and the vector $\mathbf p_n^{(t-1)}-\mathbf p_j$ is
\begin{align}\label{outhy}
	\Big\{\mathbf w\ \big|\ (\mathbf p_j-\mathbf p_{\mathrm{bdry},j})^{\mathrm T}\mathbf w=c\Big\},
\end{align}
where $c\in\mathbb R$.
The closed halfspace used for convexification is defined as
\begin{align}\label{muto0}
	\Big\{\mathbf w\ \big|\ 
	\big(\mathbf p_j-\mathbf p_n^{(t-1)}\big)^{\mathrm T}
	\big(\mathbf w-\mathbf p_{\mathrm{bdry},j}\big)\le 0
	\Big\}.
\end{align}
Setting $\mathbf w=\mathbf p_n$ yields the following linear inequality,
\begin{align}\label{bin1}
	\big(\mathbf p_j-\mathbf p_n^{(t-1)}\big)^{\mathrm T}
	\big(\mathbf p_n-\mathbf p_{\mathrm{bdry},j}\big)\le 0,
	\ \forall j\in\mathcal N\setminus\{n\}.
\end{align}
\begin{algorithm}[t]
	\caption{Conditional-Gradient-Based Coupler Position Optimization Algorithm}\label{alg:fw-cs}
	\begin{algorithmic}[1]
		\STATE \textbf{Input:} feasible set $\mathcal C$, spacing $d_{\min}$, loads $\mathbf X$, power $P$, noise $\sigma^2$, Armijo parameters $(\beta,\eta)$, tolerance $\varepsilon$, and maximum number of sweeps $T_{\max}$.
		\STATE Initialize $\{\mathbf p_n^{(0)}\}_{n=1}^N\subset\mathcal C$ with $\|\mathbf p_m^{(0)}-\mathbf p_{m'}^{(0)}\|\ge d_{\min}$.
		\FOR{$t=1,2,\ldots,T_{\max}$}
		\FOR{$n=1$ to $N$}
		\STATE Build the convex set for $\mathbf p_n$ via \eqref{eq:bdry_d}-\eqref{bin1}.
		\STATE Compute $\nabla_{\mathbf p_n}\widetilde{R}(\mathbf p_n^{(t-1)})$ using \eqref{eq:cs-grad}.
		\STATE Solve (P4) to get $\tilde{\mathbf p}_n^{(t-1)}$.
		\STATE Perform the line search \eqref{eq:armijo-new} and update by \eqref{gt}.
		\ENDFOR
		\STATE If $R(\mathbf p^{(t)})-R(\mathbf p^{(t-1)})<\varepsilon$, terminate.
		\ENDFOR
		\STATE \textbf{Output:} $\mathbf p^\star=\mathbf p^{(t)}$.
	\end{algorithmic}
\end{algorithm}

Since the feasible region of (P3) is convex, it can be solved by feasible-direction methods.
A feasible-direction method starts with a feasible vector $\mathbf p_n^{(0)}$ and generates a sequence $\{\mathbf p_n^{(t)}\}$:
\begin{align}\label{gt}
	\mathbf p_n^{(t)}=\mathbf p_n^{(t-1)}+\tau^{(t-1)}
	\big(\tilde{\mathbf p}_n^{(t-1)}-\mathbf p_n^{(t-1)}\big),
\end{align}
where $\tau^{(t-1)}\in(0,1]$ is the adaptive step size computed by the Armijo rule \cite{arm}.
We choose $\tau^{(t-1)} = \beta^{\ell}$ with $\beta, \eta \in (0, 1)$, where $\ell$ is the smallest non-negative integer satisfying
\begin{align}\label{eq:armijo-new}
	\widetilde{R}\!\big(\mathbf p_n^{(t-1)}+\beta^{\ell}
	(\mathbf p_n^{(t)}-\mathbf p_n^{(t-1)})\big)
	-\widetilde{R}(\mathbf p_n^{(t-1)}) \nonumber\\
	\ge \eta\,\beta^{\ell}\,
	\nabla_{\mathbf p_n}\widetilde{R}(\mathbf p_n^{(t-1)})^{\mathrm T}
	\big(\mathbf p_n^{(t)}-\mathbf p_n^{(t-1)}\big).
\end{align}
In \eqref{gt}, $\tilde{\mathbf p}_n^{(t-1)}$ is a feasible point different from $\mathbf p_n^{(t-1)}$, and $\tilde{\mathbf p}_n^{(t-1)}-\mathbf p_n^{(t-1)}$ is a feasible direction.
Since the feasible region is convex, the update in \eqref{gt} remains feasible.

Next, we adopt the conditional gradient method \cite{linear} to obtain $\tilde{\mathbf p}_n^{(t-1)}$.
In this method, $\tilde{\mathbf{p}}_n^{(t-1)}$ is obtained by solving
\begin{subequations}\label{eq:P4prime}
	\begin{align}
		\text{(P4)}\quad 
		&\min_{\mathbf p_n}\ 
		-\nabla_{\mathbf p_n}\widetilde{R}(\mathbf p_n^{(t-1)})^{\mathrm T}
		\big(\mathbf p_n-\mathbf p_n^{(t-1)}\big)\\
		\text{s.t. }\;
		& \mathbf p_n\in\mathcal C,\\
		& \big(\mathbf p_j-\mathbf p_n^{(t-1)}\big)^{\mathrm T}
		\big(\mathbf p_n-\mathbf p_{\mathrm{bdry},j}\big)\le 0, \forall j\neq n,
	\end{align}
\end{subequations}
where the gradient of the objective with respect to $\mathbf p_n$ at $\mathbf p_n^{(t-1)}$ is computed via the complex-step approximation \cite{css, shao2026flexible}, which retains machine-precision accuracy without subtractive cancellation.
Specifically, for $\mathbf{p}_n = [x_n, y_n, z_n]^{\mathrm T}$ with $z_n = 0$, and with a sufficiently small step size $h_{\mathrm{cs}}$, the partial derivatives with respect to the in-plane coordinates are approximated as
\begin{subequations}\label{eq:cs-grad}
	\begin{align}
		\big[\nabla_{\mathbf p_n}\widetilde{R}(\mathbf p_n)\big]_x
		&\approx 
		\frac{\Im\!\left\{\widetilde{R}\big(\mathbf p_n + \mathrm j h_{\mathrm{cs}}\,[1,0,0]^{\mathrm T}\big)\right\}}{h_{\mathrm{cs}}},\\
		\big[\nabla_{\mathbf p_n}\widetilde{R}(\mathbf p_n)\big]_y
		&\approx 
		\frac{\Im\!\left\{\widetilde{R}\big(\mathbf p_n + \mathrm j h_{\mathrm{cs}}\,[0,1,0]^{\mathrm T}\big)\right\}}{h_{\mathrm{cs}}}.
	\end{align}
\end{subequations}
Since the couplers are restricted to move in the $x$-$y$ plane, the $z$-coordinate remains fixed and is not optimized.

We sweep $n=1,\ldots,N$ and repeat \eqref{gt}-\eqref{eq:cs-grad} until the rate increment is below a tolerance or a maximum sweep count is reached.
The details of the conditional gradient algorithm for solving problem (P2) are presented in Algorithm 1.

In the following, we analyze the computational complexity of the proposed algorithm. 
The dominant computational cost arises from assembling the impedance matrices and factorizing 
$(\mathbf{Z}_{\mathrm{E}}(\mathbf{p})+\mathbf{X})$, which requires $\mathcal{O}(N^{3})$ operations. 
The complex-step gradient for each coupler is computed with a cached matrix factorization, 
which results in a complexity of $\mathcal{O}(N^{2})$. 
Hence, with $N$ couplers, $T$ outer sweeps, and an average of $S$ backtracking trials per update, 
the overall computational complexity of the proposed optimization in Algorithm~1 is 
$\mathcal{O}(TSN^{3})$.

\section{Simulation Results}\label{sec:sim}
\begin{table}[t]
	\centering
	\caption{Simulation Parameters}
	\label{tab:sim_params}
	\begin{tabular}{@{}c l@{\hspace{0.4em}} c@{}}
		\toprule
		\textbf{Symbol} & \textbf{Description} & \textbf{Value} \\
		\midrule
		$\phi$ & User direction & $30^{\circ}$ \\
		$A$ & Movement-region side length  & $0.8 \lambda$ \\
		$f_c$ & Carrier frequency & $7~\mathrm{GHz}$ \\
		$\lambda$ & Wavelength & $0.043~\mathrm{m}$ \\
		$N$ & Number of couplers & $3$ \\
		$d_{\mathrm{TU}}$ & Transmitter-user distance & $250~\mathrm{m}$ \\
		$d_{\min}$ & Minimum element spacing & $0.1\lambda$ \\
		$P_{\max}$ &Maximum transmit power & 30 dBm\\
		$L$ &Number of paths & 6\\
		\bottomrule
	\end{tabular}
\end{table}

In this section, we provide numerical results to evaluate
the performance of our proposed FCA-enhanced system design for maximizing the achievable rate. The coupler movement region is set as a square region with side length $A$.
The main
simulation parameters are provided in Table II, unless specified
otherwise. In the simulation, we model the active antenna (port index $0$) and the $N$ passive couplers (port indices $1,\ldots,N$) as thin straight wires, all parallel to the $z$-axis, with identical length $D$. The in-plane center-to-center distance between ports $m$ and $n$ is given by
\begin{align}
	d_{mn} \triangleq \|\mathbf p_m-\mathbf p_n\|_2. \label{eq:dxy}
\end{align}

The entries of the transmit impedance matrix $\mathbf Z_{\mathrm{TX}}(\mathbf{p})$ depend on the coupler positions. Specifically, we have
\begin{align}
	[\mathbf Z_{\mathrm{TX}}]_{m,n}\!=\!
	\begin{cases}
		z_{\text{self}}, & m=n\\
		z_{\text{mut}}(\mathbf p_m,\mathbf p_n), & m\neq n
	\end{cases}
	\label{eq:ZT-structure}
\end{align}
with $m,n\in\{0,1,\ldots,N\}$ representing the active antenna index $0$ and the $N$ couplers indices $1\ldots N$. 

For two side-by-side parallel identical couplers, each with length of $\lambda/2$, their mutual impedance can be written as \cite{balanis2016antenna}
\begin{subequations}\label{9o}
	\begin{align}
		\Re\{z_{\mathrm{mut}}(\mathbf p_m,\mathbf p_n)\}
		&= \frac{\eta}{4\pi}\!\left[2\,\Ci(u_0)-\Ci(u_1)-\Ci(u_2)\right],\\
		\Im\{z_{\mathrm{mut}}(\mathbf p_m,\mathbf p_n)\}
		&= -\frac{\eta}{4\pi}\!\left[2\,\Si(u_0)-\Si(u_1)-\Si(u_2)\right].
	\end{align}
\end{subequations}
where $u_0 = k\,d_{mn}$, $u_1 = k\big(\sqrt{d_{mn}^2+D^2}+D\big)$,
$u_2 = k\big(\sqrt{d_{mn}^2+D^2}-D\big)$, $S_{i}(x) = \int_{0}^{x} \frac{\sin(\tau)}{\tau} \, d\tau$, $C_{i}(x)= \int_{\infty}^{x} \frac{\cos(\tau)}{\tau} \, d\tau$,
$k=2\pi/\lambda$, and $\eta$ is the intrinsic wave impedance of the medium, which in free space is $\eta = 120\pi~\Omega$.

On the other hand, the self impedance depends only on the element geometry (e.g., coupler length and radius) and is independent of the coupler positions $\mathbf p$. According to \cite{balanis2016antenna}, its closed-form expression is given by
\begin{subequations}
	\begin{align}
		&\Re\{z_{\mathrm{self}}\}
		\!=	\!\frac{\eta}{2\pi}\Big[
		C+\ln(kD)-\mathrm{Ci}(kD)
		+\tfrac{1}{2}\sin(kD)(\mathrm{Si}(2kD) \notag\\
		&
		-2\,\mathrm{Si}(kD))+\tfrac{1}{2}\cos(kD)\big(C+\ln(kD/2)+\mathrm{Ci}(2kD) \nonumber\\
		&-2\,\mathrm{Ci}(kD)\big)
		\Big],\\
		&\Im\{z_{\mathrm{self}}\}
		= \frac{\eta}{4\pi}\Big\{
		2\,\mathrm{Si}(kD)
		+\cos(kD)\big[2\,\mathrm{Si}(kD)-\mathrm{Si}(2kD)\big] \notag\\
		&
		-\sin(kD)\big[2\,\mathrm{Ci}(kD)-\mathrm{Ci}(2kD)-\mathrm{Ci}\!\left(2ka^{2}/D\right)\big]
		\Big\}, 
	\end{align}
\end{subequations}
where $C\approx0.5772$ is the Euler-Mascheroni constant, $a$ denotes the radius of the thin straight wire dipole in free space. In this paper, we adopt thin-wire dipoles with approximately omnidirectional azimuth-plane patterns. In practice, directional antennas with higher element gains can improve link robustness and broaden the applicability of the proposed architecture.

For comparison, the proposed FCA is compared with the following baseline schemes.
\begin{itemize}
\item \textbf{Single active antenna}: 
The transmitter is equipped with a single active antenna without any couplers.
\item \textbf{Fixed-position coupler}: 
The $N$ couplers are fixed at uniformly spaced positions with spacing $0.4\lambda$ around the active antenna within the predefined array geometry, and the load impedance matrix $\mathbf{X}$ is fixed as a diagonal matrix with diagonal entries equal to $0.05 + \mathrm{j}50~\Omega$.
\item \textbf{ESPAR}: 
The load impedance matrix $\mathbf{X}$ is optimized in closed form \cite{ESP}, while the $N$ coupler positions are fixed to be the same as those in the fixed-position coupler case.
\item \textbf{Active antenna array}: 
All $N+1$ ports are configured as active antennas and arranged as a fixed uniform planar array (UPA) with half-wavelength inter-element spacing.
\item \textbf{Translatable 6DMA}: 
All $N+1$ ports are configured as active antennas and can move within regions of the same size as the coupler movement region.
\end{itemize}

\subsection{LoS Channel}
\begin{figure}[t!]
	\centering
	\setlength{\abovecaptionskip}{0cm}
	\subfloat[Configuration 1.
	\label{fig:CST1}]{
		\includegraphics[width=1.58in]{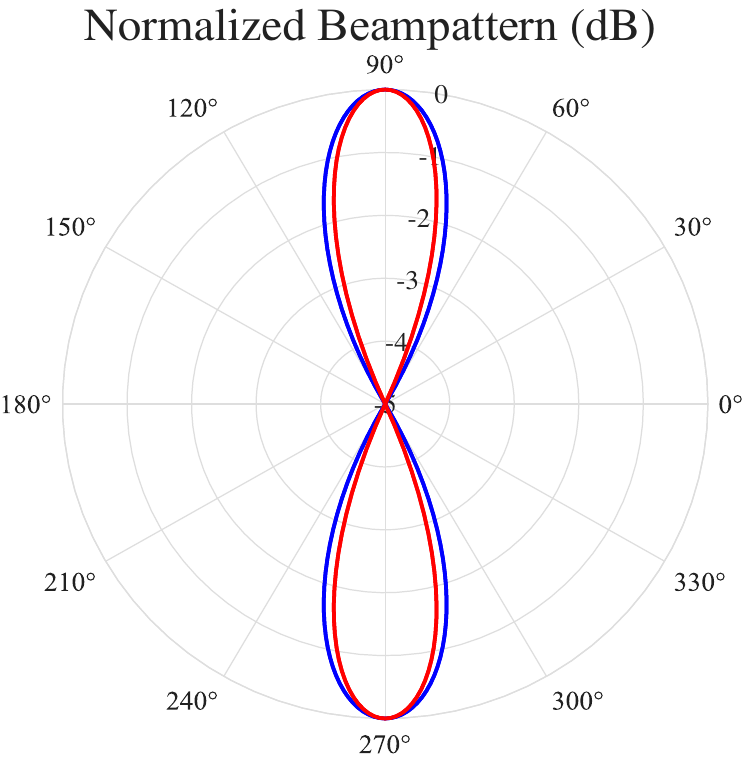}
	}
	\hspace{-0.07in}
	\subfloat[Configuration 2.
	\label{fig:CST2}]{
		\includegraphics[width=1.75in]{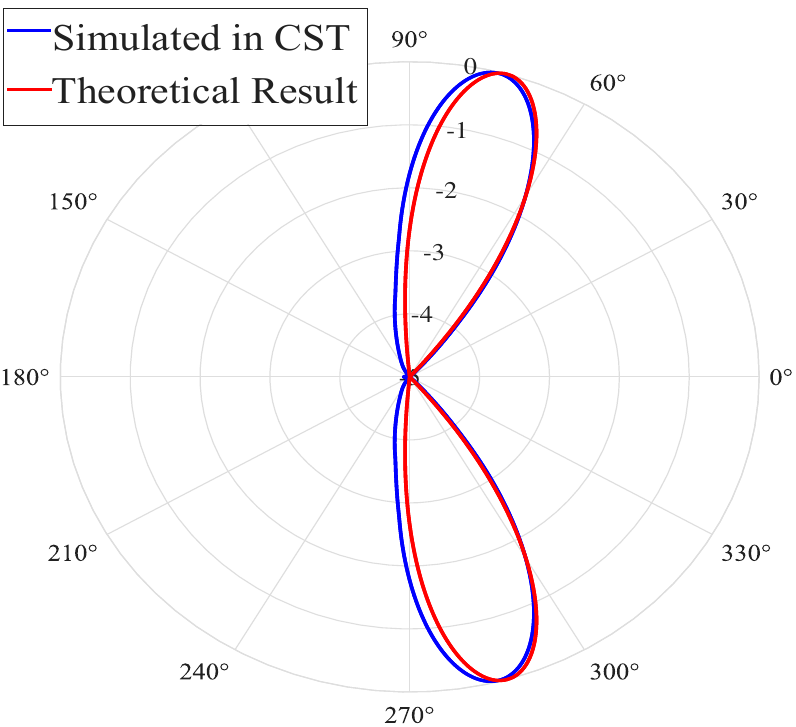}}
	\caption{Comparison of the normalized beampatterns obtained by CST simulation and theoretical analysis.}
	\label{fig:CST}
\end{figure}

In Fig. \ref{fig:CST}, we validate the theoretical expression of the FCA SNR in \eqref{buu1} for the LoS channel based on the circuit theory approach through computational electromagnetic simulations in computer simulation technology (CST) Studio Suite. 
We consider one active antenna located at the origin and two passive couplers positioned along the x-axis in the $x$-$y$ plane. 
In CST, the active antenna at \((0,0)\) is excited by a voltage source with unit magnitude and zero phase, while the two passive couplers are modeled as thin-wire dipoles terminated with the same fixed load impedance of \(0.05+\mathrm{j}50~\Omega\). Two representative coupler-position configurations are considered, namely, coupler 1 and coupler 2 located at \(({-0.6}\lambda,0)\) and \((0.6\lambda,0)\), respectively, for configuration 1, and at \(({-0.5}\lambda,0)\) and \((0.8\lambda,0)\), respectively, for configuration 2. 
The normalized beampatterns in the azimuth plane for the two configurations are shown in Fig.~\ref{fig:CST}(a) and Fig.~\ref{fig:CST}(b). In both cases, the CST-simulated and theoretical beampatterns are well aligned, which supports the validity of the adopted circuit-theory approach for the considered dipole-type FCA configuration.

\begin{figure}[t!]
	\centering
	\setlength{\abovecaptionskip}{0.cm}
	\includegraphics[width=3.56in]{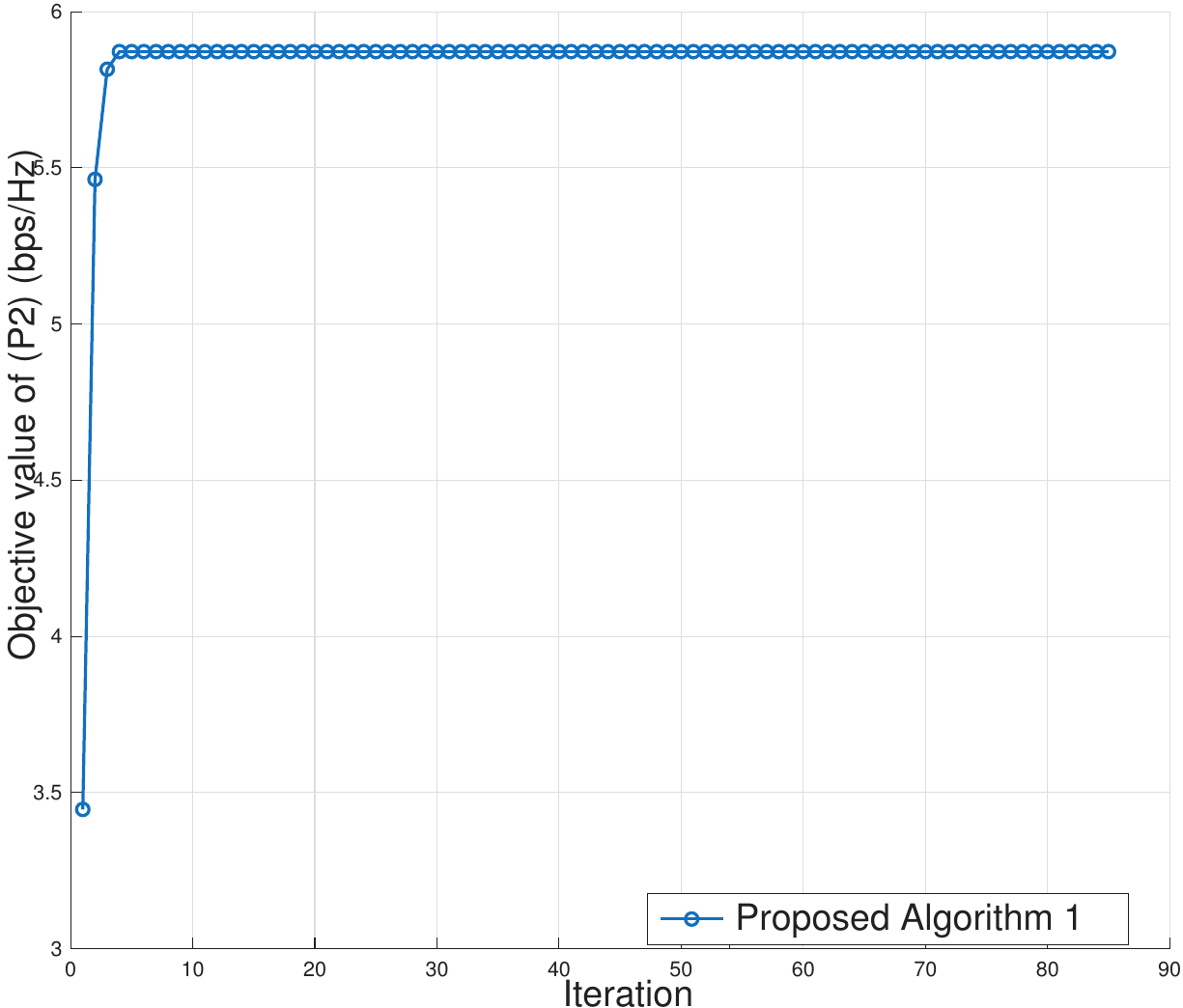}
	\caption{Convergence of the flexible-coupler position optimization algorithm.}
	\label{convergence}
\end{figure}

For the LoS channel (i.e., $L=1$), Fig.~\ref{convergence} illustrates the convergence behavior of the proposed Algorithm 1 for the FCA-enabled system. As can be observed, starting from a feasible initialization, the achievable rate increases monotonically with the iteration index under the conditional-gradient updates, and it stabilizes after about 5 iterations. Subsequent iterations yield negligible improvement, which verifies that the block-coordinate scheme reaches a stationary solution quickly for the adopted parameters. 

\begin{figure}[t!]
	\centering
	\setlength{\abovecaptionskip}{0.cm}
	\includegraphics[width=3.59in]{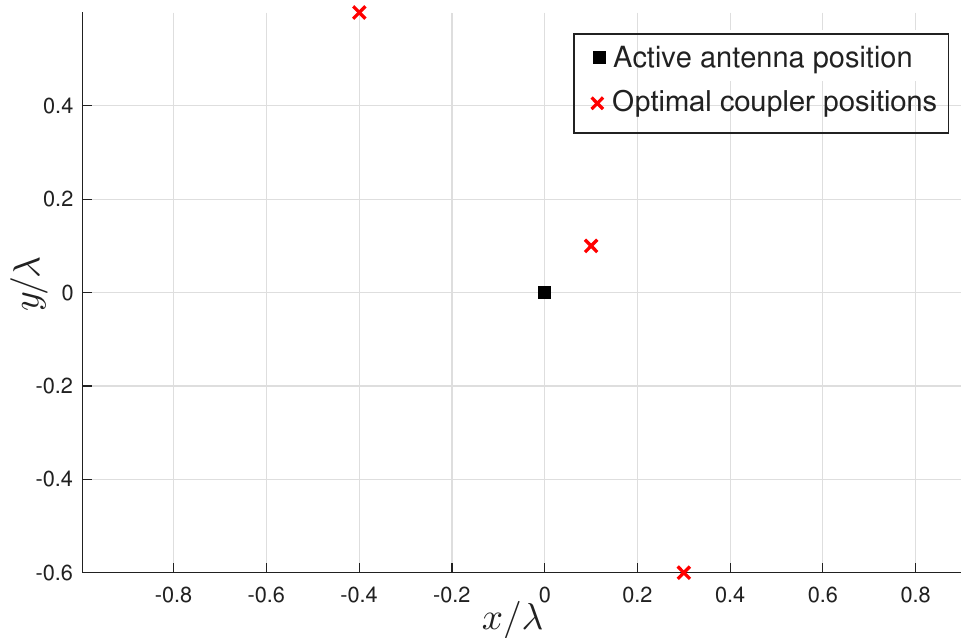}
	\caption{Optimized coupler positions around the active antenna. }
	\label{sight}
\end{figure}

\begin{figure}[t!]
	\centering
	\setlength{\abovecaptionskip}{0.cm}
	\includegraphics[width=3.56in]{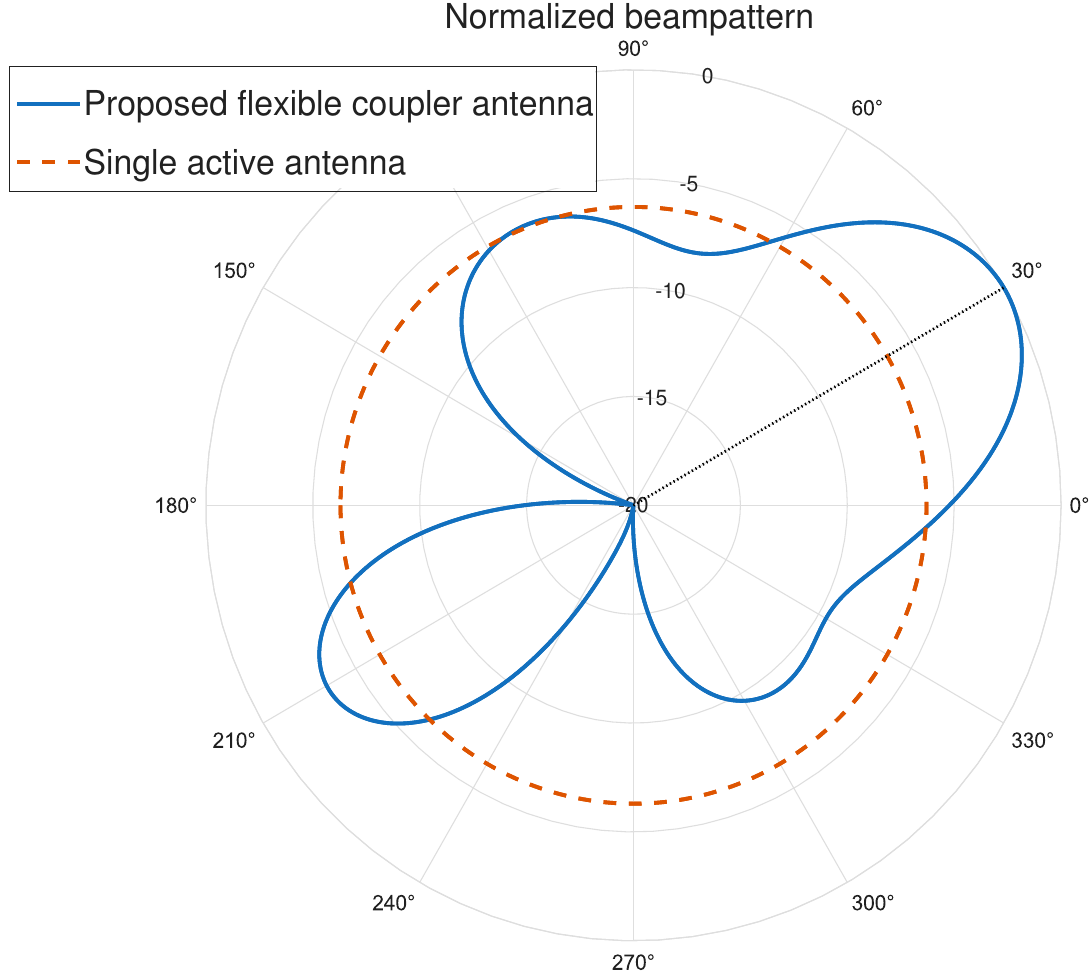}
	\caption{The normalized beampattern (dB) at user direction $\phi=30^{\circ}$.}
	\label{patt}
\end{figure}

In Fig.~\ref{sight}, we plot the numerically optimized positions
$\{\mathbf p_n^\star\}_{n=1}^{N}$ of the couplers around the fixed active antenna at the origin. All passive couplers lie in the $x$-$y$ plane within the feasible region
$\mathcal C$ with $z_n=0$. The optimized geometry concentrates the couplers on the side facing
the user receiver at $\phi=30^{\circ}$, with inter-element distances no less than $d_{\min}$ to enlarge
the effective aperture while exploiting beneficial mutual coupling. This arrangement strengthens
the equivalent weight vector $\tilde{\mathbf w}(\mathbf p^\star)$ and steers the main lobe
toward the user direction, which is consistent with the normalized beam pattern subsequently shown in Fig.~\ref{patt}.

Substituting the optimized coupler positions into \eqref{buu1}, we obtain the transmit beampattern of the proposed FCA. Fig.~\ref{patt} compares the transmit beampatterns of the proposed FCA and the single-active-antenna baseline. Fig.~\ref{patt} shows that the main lobe of the proposed FCA aligns with the user direction \(\phi=30^\circ\), which indicates that currents induced by mechanically translating the coupler positions steer energy toward the target direction. Compared with the single-active-antenna scheme, the proposed FCA exhibits a more concentrated beam pattern toward the user direction and improved angular selectivity, which indicates that the position-dependent mutual impedance shapes the equivalent weights in \(\tilde{\mathbf w}(\mathbf p)\) to achieve the desired beam pattern. These results demonstrate that the proposed FCA can enhance directional transmission without requiring multiple active antennas.

\begin{figure}[t!]
	\centering
	\setlength{\abovecaptionskip}{0.cm}
	\includegraphics[width=3.56in]{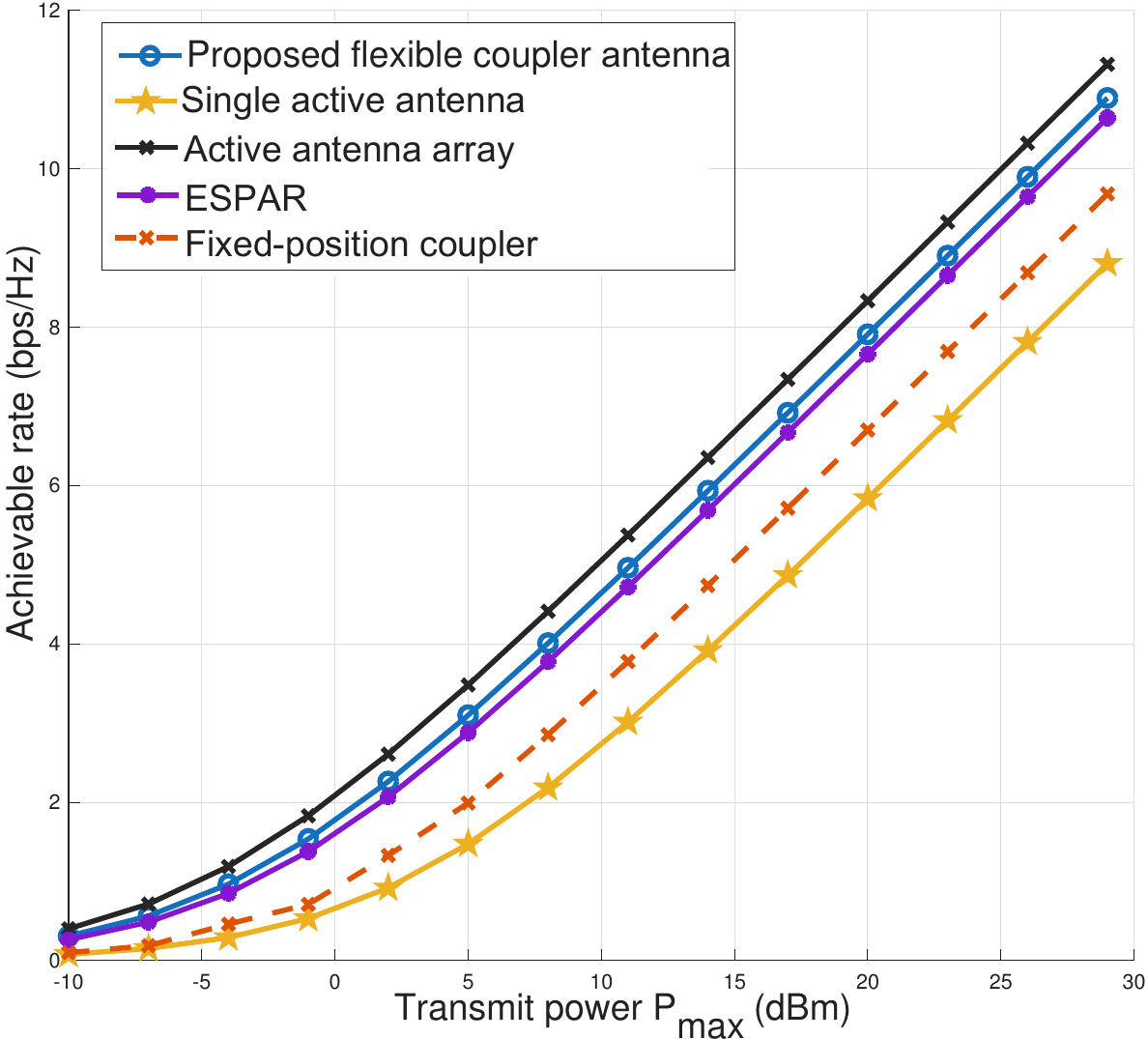}
	\caption{Rate versus maximum transmit power for different schemes under the LoS channel.}
	\label{px}
\end{figure}

Using \eqref{buu1}, the achievable rate of the user in bits-per-second-per-Hertz (bps/Hz) is given by
\begin{align}\label{bbuu}
	R(\mathbf p)=\log_2\!\big(1+r(\mathbf p)\big).
\end{align}
Fig.~\ref{px} plots the achievable rate versus the maximum transmit power \(P_{\max}\) under the LoS channel. The proposed FCA-enhanced system  exhibits a substantial rate gain over the fixed-position coupler and single active antenna baselines, and a small gain over the ESPAR.
The improvement over the fixed-position coupler arises from exploiting the new position adjustment DoF to reshape the position-dependent mutual impedance, which strengthens the effective channel gain and array directivity toward the user direction. Since the baseline antenna/array  geometry is not optimized, the rate gap enlarges as \(P_{\max}\) increases. 
Compared with the ESPAR that relies on load impedance optimization with fixed geometry, mechanically translating passive couplers provides a new flexible  control of the induced currents without the need for high-precision tunable hardware components. This leads to higher rate at lower circuit  cost and complexity, which yields a more favorable performance-cost-complexity balance. 
Relative to the single active antenna, the FCA benefits from reconfigurable induced currents on the moving passive elements. Moreover, the coherent signal combination of the single active antenna and passive couplers increases the effective aperture and improves the rate without additional RF chains or extra circuit power beyond the single RF chain.
The active antenna array attains the highest rate because each element is separately controllable based on the principle of maximum ratio transmission (MRT). The FCA approaches this benchmark closely and requires about \(0.4\,\mathrm{dB}\) more transmit power to reach the same rate. Despite this small power offset, the FCA attains the highest energy efficiency among all schemes considered, since it avoids the large circuit-power overhead inherent to fully active arrays while still offering  strong beamforming gains.

\begin{figure}[t!]
	\centering
	\setlength{\abovecaptionskip}{0.cm}
	\includegraphics[width=3.5in]{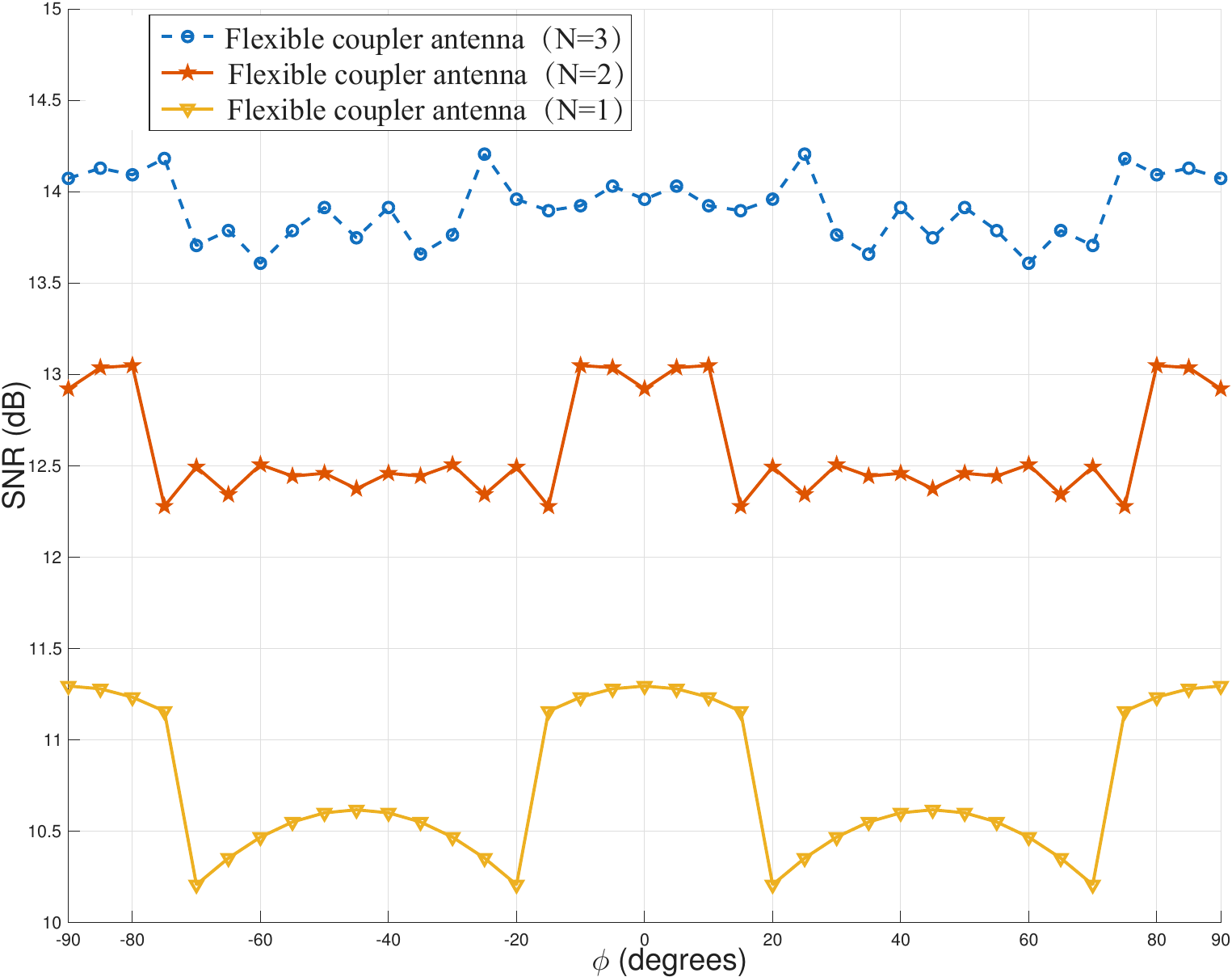}
	\caption{Received SNR versus angle for different values of $N$.}
	\label{degree}
\end{figure}

In Fig.~\ref{degree}, we plot the received SNR versus the user angle for different numbers of passive couplers. For each angle, the coupler positions are optimized within the admissible movement region.  An interesting property of the FCA is that the SNR increases with the number of passive couplers $N$, without the need to add more active antennas.
The variation of the maximum SNR across angles decreases with larger \(N\), which indicates higher robustness of the mechanically reconfigurable architecture to the user direction. 
Unlike the superdirectivity effect observed at the endfire angle for a fixed-position array \cite{10104538}, the proposed FCA design does not necessarily favor endfire over broadside, because the geometry is reconfigured at each angle and the endfire superdirectivity effect of fixed arrays no longer dominates.
Moreover, the received SNR is symmetric with respect to the user angle, since the optimal geometry for $-\phi$ is the mirror image of that for $+\phi$ with the active element located at the origin, yielding an even response in $\phi$.

\begin{figure}[t!]
	\centering
	\setlength{\abovecaptionskip}{0.cm}
	\includegraphics[width=3.5in]{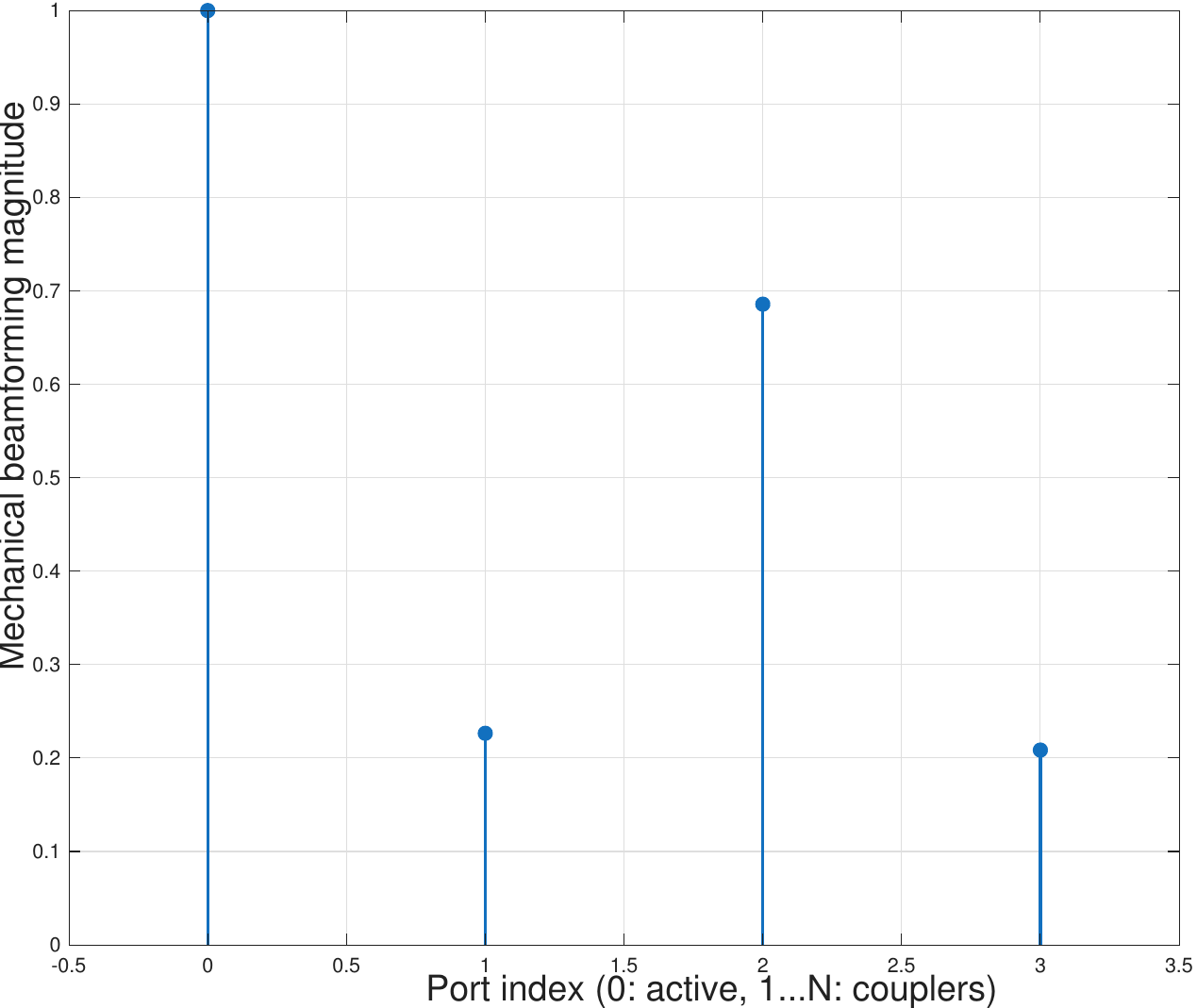}
	\caption{Magnitudes of mechanical beamforming weights after coupler position optimization.}
	\label{mag}
\end{figure}
\begin{figure}[t!]
	\centering
	\setlength{\abovecaptionskip}{0.cm}
	\includegraphics[width=3.48in]{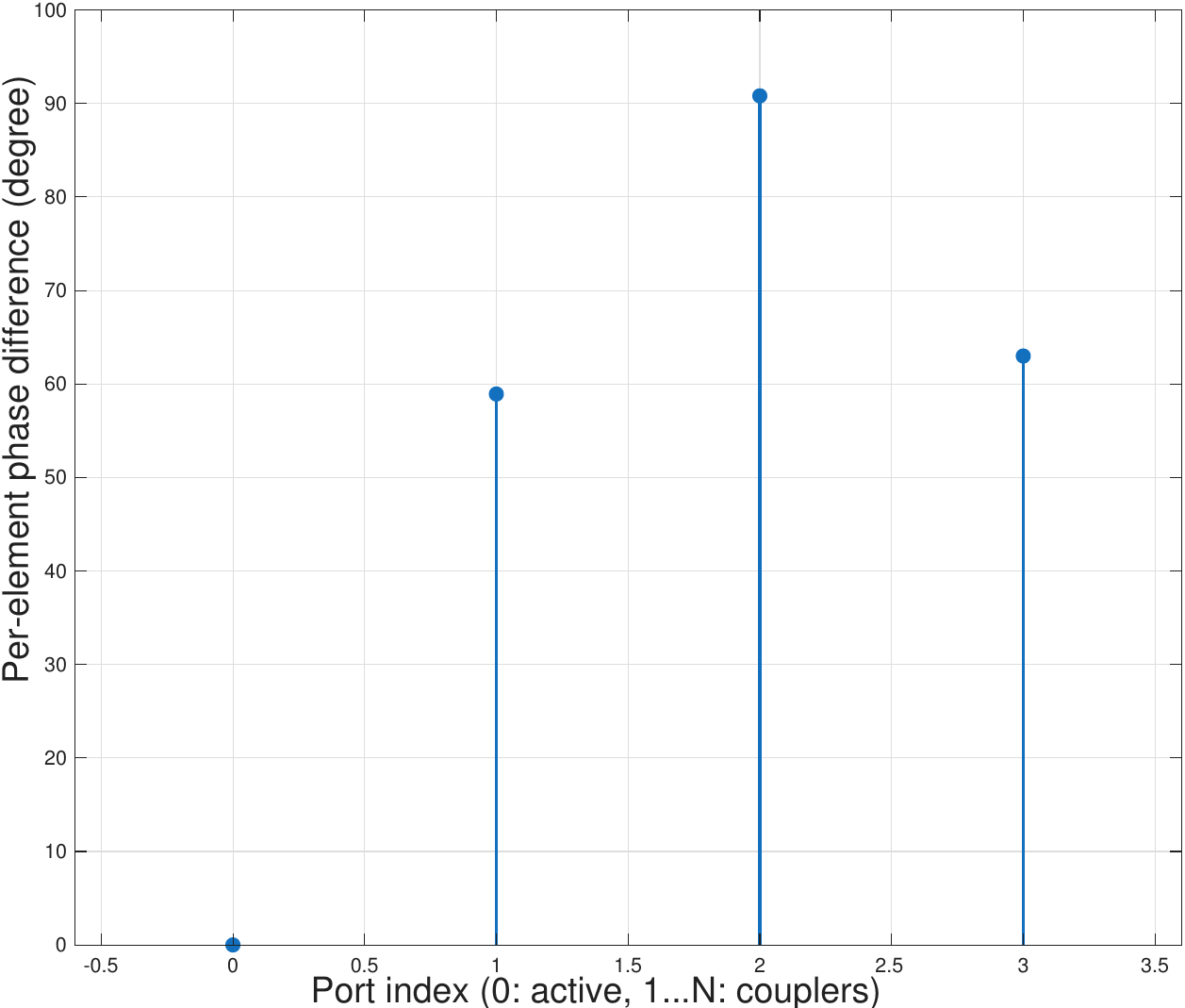}
	\caption{Per-element phase difference after coupler position optimization.}
	\label{phase}
\end{figure}

Fig. \ref{mag} shows the magnitudes ($\left|[\tilde{\mathbf{w}}]_i\right|,\ i=1,\ldots,N$
) after coupler position optimization for the FCA transmitter. Port index 0 denotes the active antenna, and indices 1 to $N$ correspond to the passive couplers. The active port exhibits the largest magnitude, while only coupler 2  attains significant magnitude, and those of the other two couplers (1 and 3)  are suppressed to prevent destructive interaction with the main lobe, due to phase misalignments between the mechanical beamforming weights in $\tilde{\mathbf{w}}(\mathbf p)$ and the corresponding channel responses 
in $\mathbf h(\mathbf p)$. The optimized magnitudes of beamforming weights thus enhance the coherent signal combination $| \mathbf h^{\mathrm T}(\mathbf p)\tilde{\mathbf w}(\mathbf p)|$, which explains the rate improvement observed over the fixed-position and conventional ESPAR baselines.
In Fig.~\ref{phase}, we plot the angle differences between the mechanical  beamforming weights and corresponding channel responses, i.e., $ \angle [\tilde{\mathbf{w}}]_i - \angle [\mathbf{h}]_i$, after coupler position optimization, which represent the undesired phase misalignments between them. The active port $i=0$ shows a near-zero phase difference, indicating that its phase aligns with the corresponding channel response. The passive coupler ports exhibit residual phase differences due to coupled magnitude and phase constraints. 
From Figs.~\ref{mag} and~\ref{phase}, it can be observed that through coupler position optimization, the proposed algorithm jointly tunes the magnitudes and phases of the beamforming weights so that the overall induced currents become approximately conjugate-aligned with the user channel responses, thereby maximizing the mechanical beamforming gain.
\subsection{Multipath Channel}
\begin{figure}[t!]
	\centering
	\setlength{\abovecaptionskip}{0.cm}
	\includegraphics[width=3.5in]{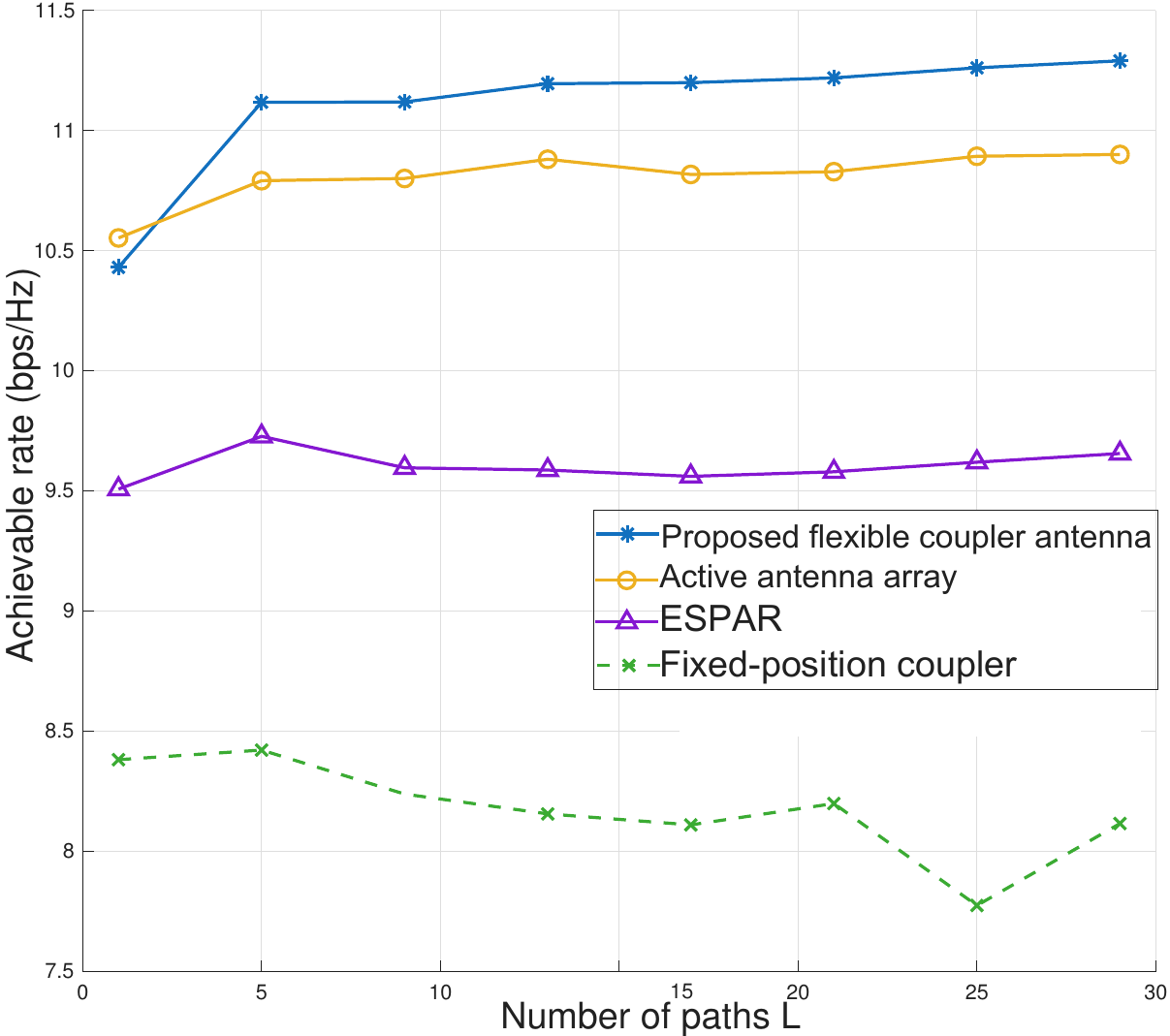}
	\caption{Achievable rates of different schemes versus the
		number of channel paths.}
	\label{path}
\end{figure}

For the case of multipath channel, Fig. \ref{path} shows the achievable rates of different schemes versus the number of channel paths $L$. 
As \(L\) increases, the proposed FCA scheme not only improves the achievable rate steadily but also outperforms the fully active array. 
This result is due to the combined mechanical beamforming and spatial diversity gains obtained through the coupler position optimization based on the channel distribution in their movable region. Specifically, the position optimization reshapes the position-dependent mutual impedance and enables mechanical beamforming, which strengthens the effective channel gain \(|\mathbf h^{\mathsf T}(\mathbf p)\tilde{\mathbf w}(\mathbf p)|\) that determines the received SNR and achievable rate. 
Moreover, by relocating the passive couplers within the feasible region, the transmitter mitigates deep channel fading of multiple paths over the feasible region by exploiting the spatial diversity from multipath superposition. 
These benefits jointly enhance the received SNR or achievable rate more significantly as compared to the previous LoS channel case. 
In contrast, the fully active array cannot mitigate the channel deep fading by relocating the antenna elements to avoid deep fading locations, because they remain fixed at the transmitter. 
Consequently, in rich multipath conditions, the FCA achieves an additional fading-mitigation gain in addition to the mechanical beamforming gain, which results in higher achievable rates than the fully active array even with a single RF chain. Furthermore, it is observed that when \(L\) exceeds five, the additional rate improvement with increasing $L$ becomes negligible, because the fading-mitigation gain achieved through coupler position adjustment becomes saturated when the number of channel paths is sufficiently large. 

\begin{figure}[t!]
	\centering
	\setlength{\abovecaptionskip}{0.cm}
	\includegraphics[width=3.5in]{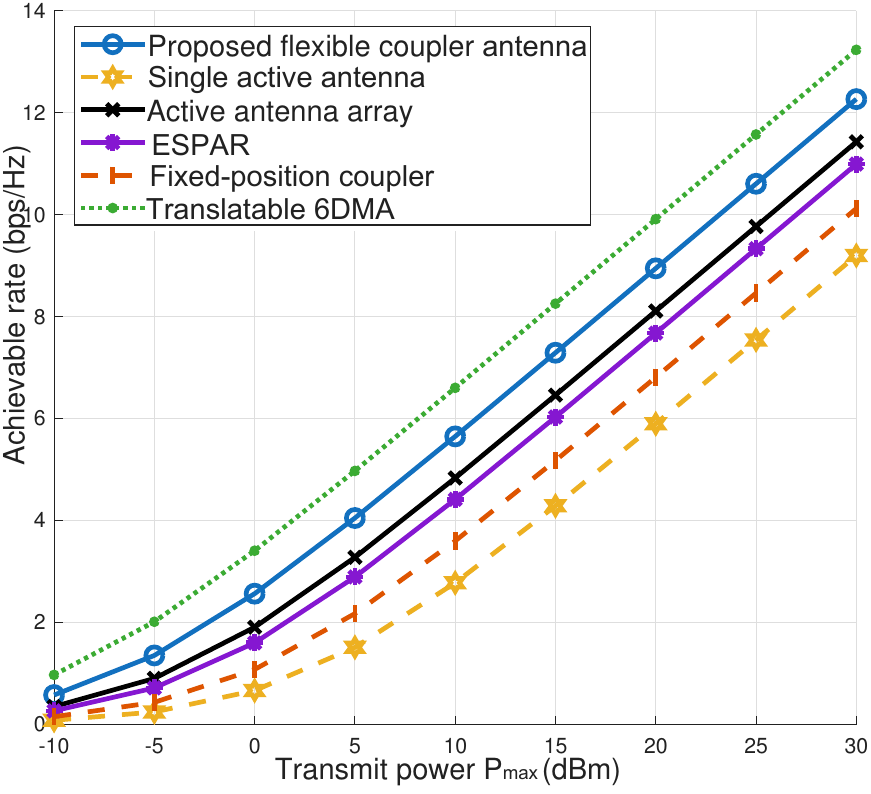}
	\caption{Rate versus maximum transmit power for different schemes under the multipath channel.}
	\label{nlospower}
\end{figure}

In Fig.~\ref{nlospower}, we plot the achievable rate versus the maximum transmit power $P_{\max}$ under the multipath channel with $A=1.3\lambda$. The figure confirms that the proposed FCA scheme outperforms the fixed-position coupler, the single active antenna, and the ESPAR across the entire power range. Although the translatable 6DMA achieves the highest rate among all schemes \cite{IPA,near,jiang2026multi}, the proposed FCA remains close to its performance while requiring only one RF chain and only low-power local MEMS actuation. By contrast, the single active antenna, the fixed-position coupler, and the ESPAR involve lower hardware complexity but achieve lower rates due to the lack of effective position reconfiguration gain. Therefore, the proposed FCA provides an effective tradeoff among communication performance, hardware cost, and energy efficiency.

\begin{figure}[t!]
	\centering
	\setlength{\abovecaptionskip}{0.cm}
	\includegraphics[width=3.56in]{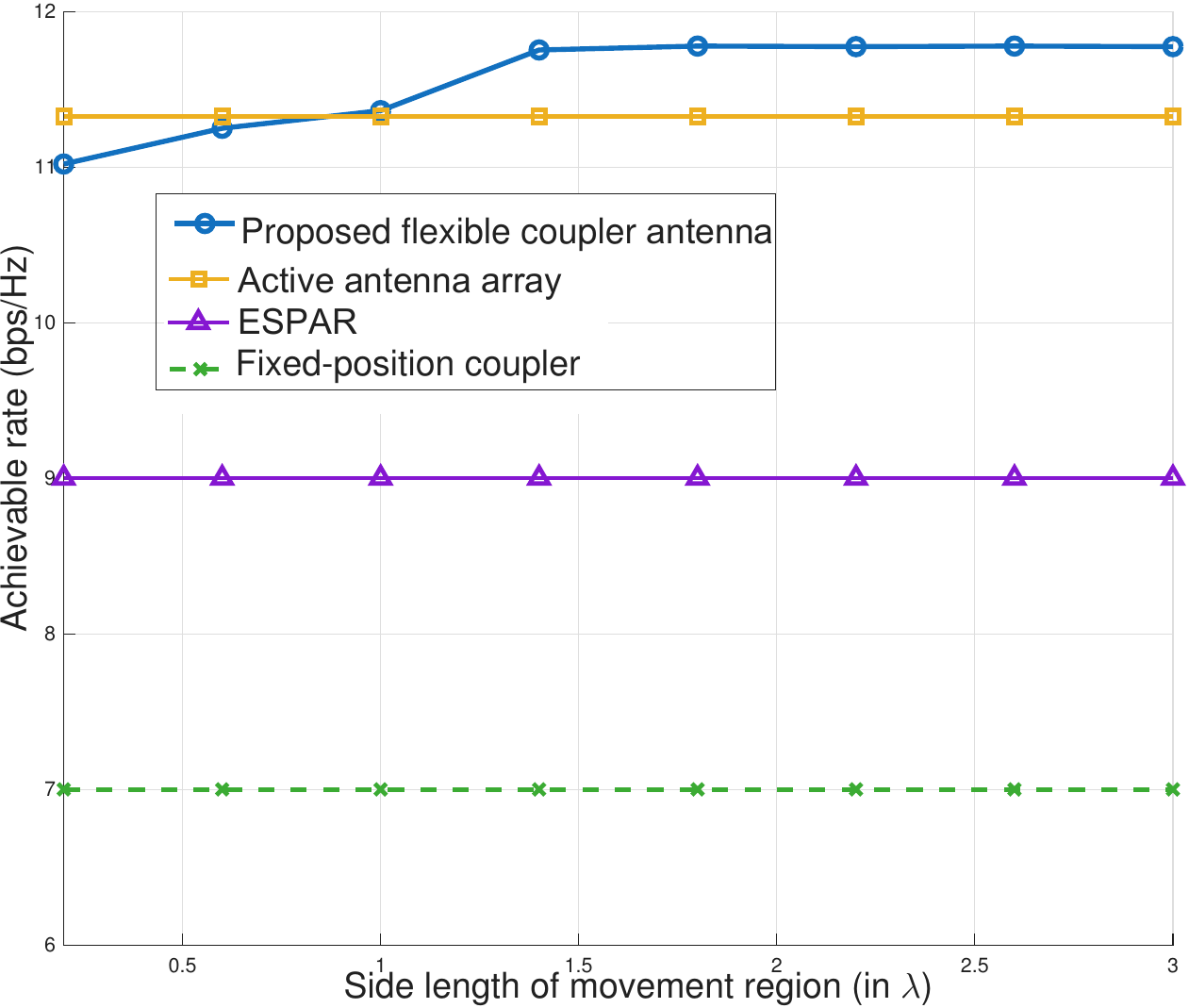}
	\caption{Achievable rate versus movement region size.}
	\label{region}
\end{figure}

In Fig.~\ref{region}, we plot the achievable rate versus the movement-region side length \(A\) for the proposed FCA system and the benchmark schemes. The proposed FCA system outperforms the ESPAR and the fixed-position coupler, and its achievable rate increases with \(A\). The rate curve of the FCA system saturates when $A$ reaches around  \(1.5\lambda\), which indicates that the maximum rate of the FCA system can be achieved with finite movement regions at the transmitter. For small movement regions with \(A\) less than $0.7\lambda$, the proposed scheme performs slightly below the active antenna array because the available translational DoF is limited and the effective aperture remains small. Moreover, when the movement region is too small, the inter-coupler distances become nearly identical and the mutual-impedance structure lacks diversity, which makes \((\mathbf Z_{\mathrm{E}}(\mathbf p)+\mathbf X)\) in \eqref{hhh} ill-conditioned. Its inversion is then numerically unstable, and the induced-current shaping is ineffective, further limiting the achievable rate. As \(A\) grows, coupler position optimization more effectively reshapes the position-dependent mutual impedance, mitigates location-dependent deep fading, and thereby enlarges mechanical beamforming gains. Its achievable rate thus surpasses that of the active array when $A$ is approximately larger than $\lambda$. The benchmark curves without mechanical reconfiguration are nearly flat with respect to \(A\) because their antenna/array geometries are fixed and no additional spatial DoF can be exploited for performance enhancement in multipath fading channels.

\begin{figure}[t!]
	\centering
	\setlength{\abovecaptionskip}{0.cm}
	\includegraphics[width=3.53in]{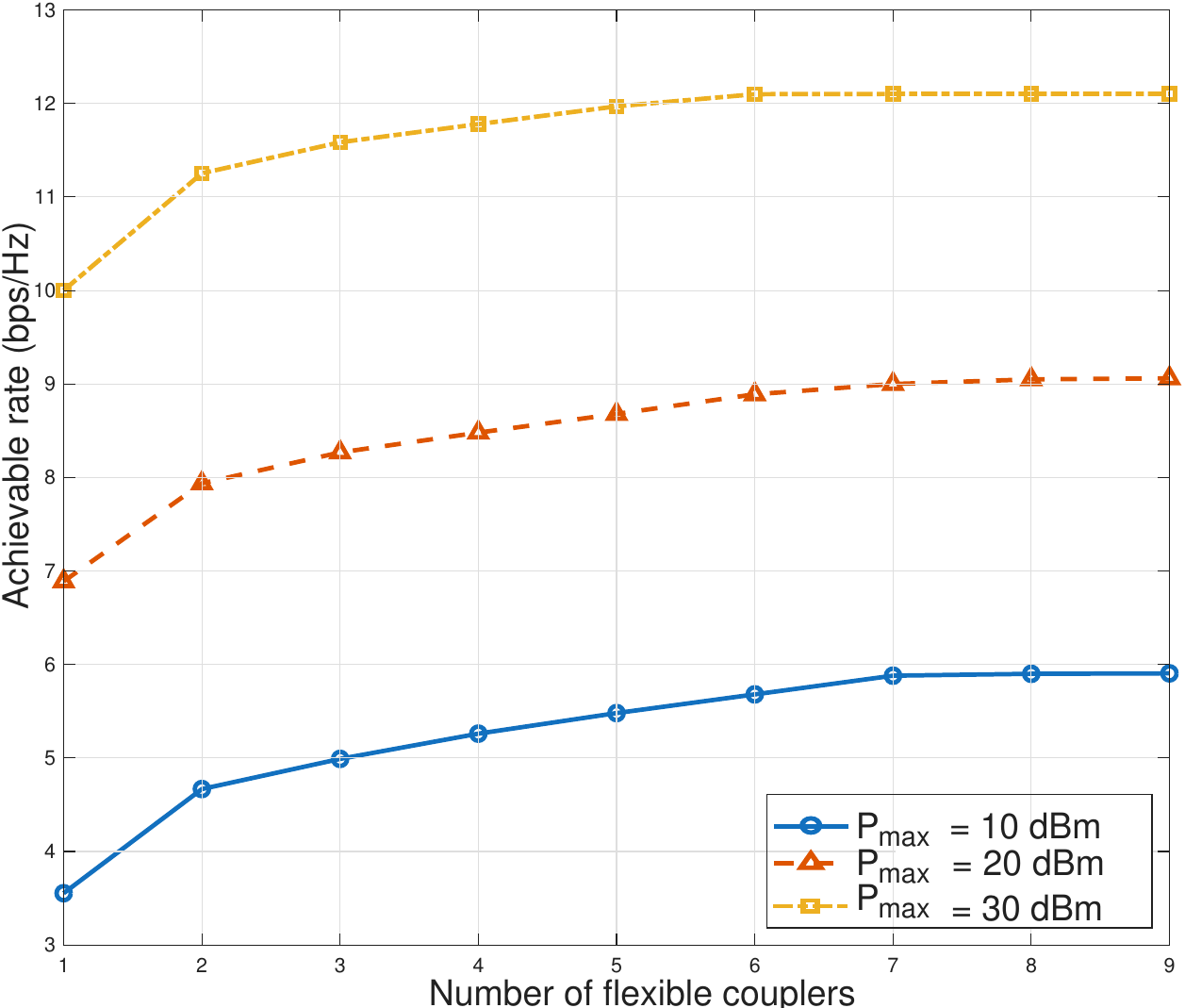}
	\caption{Achievable rate versus the number of flexible couplers.}
	\label{N}
\end{figure}

Finally, in Fig.~\ref{N}, we plot the achievable rate as a function of the number of flexible couplers \(N\). 
The rate increases up to \(N=7\) and then saturates, which indicates diminishing returns when adding more passive elements around the active antenna. 
This saturation arises for three reasons. 
First, within a finite movement region, adding more couplers creates many element pairs with similar separations, which yields highly correlated steering vectors and contributes little new spatial DoF, thus the incremental increase in \(\lvert \mathbf h^{\mathsf T}(\mathbf p)\tilde{\mathbf w}(\mathbf p)\rvert\) is marginal. 
Second, under a fixed minimum spacing, the rows and columns of the position-dependent mutual-impedance matrix \(\mathbf Z_{\mathrm E}(\mathbf p)\) become strongly correlated as \(N\) grows, so \(\mathbf Z_{\mathrm E}(\mathbf p)\) becomes increasingly ill-conditioned. 
The feasible set of passively induced currents is then restricted, which limits the effective beamforming gain. 
Third, amplitudes and phases at passive ports cannot be tuned independently, so additional couplers cannot be fully exploited for performance enhancement.
\section{Conclusions}
In this paper, we proposed a novel energy-efficient and low-cost antenna architecture called the FCA. Specifically, 
by exploiting the mechanical movement of passive couplers and the tunable mutual coupling between the couplers and active antenna, the proposed design enables the passive couplers to radiate effectively through the excitation induced by the active antenna, thereby realizing mechanical beamforming without moving any active antenna and RF front end. Given the coupler movement and 
transmit power constraints, the coupler positions were optimized
to maximize the received SNR in an FCA-enhanced single-user point-to-point communication system. By applying the block-coordinate conditional gradient 
techniques, an efficient algorithm was proposed to solve the formulated optimization problem.
Simulation results showed that, by exploiting mechanical beamforming gain in LoS channels, the FCA approaches the fully active array in achievable rate, but with significantly reduced numbers of active antennas and associated RF chains. Moreover, by jointly leveraging mechanical beamforming gain and fading-mitigation gain in multipath channels, the proposed architecture can even outperform fully active antenna arrays, which validates its potential as a promising energy-efficient solution for reconfigurable wireless networks.
In future work, we will extend the proposed design to dense-network scenarios with multiple nearby transmitters and investigate channel estimation for FCA systems, which remain challenging tasks, since each new configuration of passive coupler positions results in a different channel as well as varied coupling with the active antenna element. 
\bibliographystyle{IEEEtran}
\bibliography{fabs}
\end{document}